\documentclass[10pt,aps,pre,onecolumn,superscriptaddress,floatfix,nofootinbib,notitlepage,preprintnumbers]{revtex4-1}
\usepackage{amsmath,amssymb,amsfonts}
\usepackage{graphicx,color}
\usepackage{dcolumn}
\usepackage{bm}
\usepackage{hyperref}

\newcommand{\ie}{\emph{i.e.}}
\newcommand{\eg}{\emph{e.g.}}

\begin{document}

\title{The scientific influence of nations on global scientific and technological development}

\author{Aurelio Patelli}
\affiliation{\small Istituto dei Sistemi Complessi (ISC-CNR), 00185 Rome (Italy)}
\author{Giulio Cimini}
\affiliation{\small IMT School for Advanced Studies, 55100 Lucca (Italy)}
\affiliation{\small Istituto dei Sistemi Complessi (ISC-CNR), 00185 Rome (Italy)}
\author{Emanuele Pugliese}
\affiliation{\small Istituto dei Sistemi Complessi (ISC-CNR), 00185 Rome (Italy)}
\author{Andrea Gabrielli}
\affiliation{\small Istituto dei Sistemi Complessi (ISC-CNR), 00185 Rome (Italy)}
\affiliation{\small IMT School for Advanced Studies, 55100 Lucca (Italy)}

\begin{abstract}
Determining how scientific achievements influence the subsequent process of knowledge creation 
is a fundamental step in order to build a unified ecosystem for studying the dynamics of innovation and competitiveness. 
Relying separately on data about scientific production on one side, through bibliometric indicators, and about technological advancements on the other side, through patents statistics, 
gives only a limited insight on the key interplay between science and technology which, as a matter of fact, move forward together within the innovation space. 
In this paper, using citation data of both research papers and patents, we quantify the direct influence 
of the scientific outputs of nations on further advancements in science and on the introduction of new technologies. 
Our analysis highlights the presence of geo-cultural clusters of nations with similar innovation system features, 
and unveils the heterogeneous coupled dynamics of scientific and technological advancements. 
This study represents a step forward in the buildup of an inclusive framework for knowledge creation and innovation.
\end{abstract}

\maketitle


\section{Introduction}\label{sec:Introduction}

Developing a comprehensive conceptual framework capturing the emergent properties of the knowledge creation process requires, 
as building blocks, quantitative indicators providing insights into the structure and dynamics of innovation systems.
In this respect, numerous metrics for the impact of scientific research based on publication outputs exist in the literature---see \citet{Waltman2016} for a recent overview of the field. 
Similar (yet less refined) indicators for technological development based on patent data have been introduced as well \citep{Kurtossy2004,Nagaoka2010}. 
However, the majority of these metrics focus on either scientific or technological activities separately. 
Nevertheless, any effort for a thorough understanding of the innovation system cannot leave out of consideration the interactions between scientific and technological developments. 
Indeed, all of the recent models of knowledge production---from the ``Mode 2'' model \citep{gibbons1994} and the National Innovation System view \citep{lundvall1988} 
to the Triple Helix models \citep{etzkowitz2000}---involve non-academic forces shaping the scientific process, and identify different actors and stakeholders 
of scientific production (Firms, State) mostly involved with the spillovers of scientific work on innovation and economic performances.
In this paper we focus on specific aspects of the knowledge transfer process, namely those within science and from science to technology, at the country-level. 
To this end, we propose two bibliometric indicators based on citations that journal research papers receive from other papers as well as from patents. 

Most of the standard bibliometric impact indicators for science are indeed based on the analysis of the citations received by scientific publications from other journal papers \citep{Waltman2016}. 
The underlying assumption that citations actually reflect scientific importance, and as such are appropriate to measure scientific impact, is controversial. 
Many scientometricians agree on the fact that ``citing behavior is not motivated solely by the wish to acknowledge intellectual and cognitive influences of colleague scientists, 
since the individual studies reveal also other, in part non-scientific, factors that play a part in the decision to cite'' \citep{Bornmann2008}---such as improper citation practices 
like boosting self or friend's citations, or satisfying referees~\citep{Werner2015}. Citing also appears to be understood as a non-trivial psychological process \citep{Nicolaisen2007}. 
Thus, overall there is a large consensus on the fact that citations cannot provide an ``ideal'' monitor on scientific performance at a statistically low aggregation level 
(\eg, individual researchers), but that they can yield a strong indicator of scientific performance when considered at the level of large group of researchers and over a long period of time \citep{vanRaan2005}. 
In any event, by relying only on citations within research papers, scientific impact metrics can only assess how much a given scientific achievement 
is relevant for the community of researchers, neglecting its potential influence on other research and development (R\&D) areas.

In this respect, references to research papers listed in patents as {\em prior art} can be used to assess the importance of scientific research on technology outputs \citep{Callaert2006}. 
The mainstream approach \citep{Narin1997,Narin2000} is to compute the {\em science intensity} parameter, 
namely the average number of references to scientific literature per patent. While originally intended to characterize the scientific base of a company's patent portfolio, 
this indicator has been subsequently used for discovering the value of scientific research and forecasting future disruptive technologies. 
However, whether patent citations to papers reflect a flow of knowledge from science to technology is an even more debated issue than for paper citations. 
In fact, references in the scientific literature are added by the authors, supposedly to acknowledge existing work on which the article builds (with the exceptions outlined above). 
References in patents on the other hand have a precise legal function \citep{Callaert2006}: ``they are brought by the applicant/inventor to the attention of examiners, 
who ultimately decide which references are relevant to evaluate novelty and inventiveness, to qualify the claims made in the patent, and at last to decide on granting''. 
Inventors opinions on the meaning of patent references to papers, collected in a recent survey \citep{Callaert2014}, 
suggest that while about one-third of patents that were inspired by scientific knowledge do not contain any scientific references, half of actual scientific references 
are evaluated at least as important (\ie, having directly contributed to the inventive process), and only 10\% as not important (see Table 6 therein). 
Despite other issues affecting patent citation data, such as the difference between patent offices practices \citep{Nelson2009}, 
these and others observations suggest that patent citations to papers can be considered an indicator of the relevance of scientific findings 
for assessing and contextualizing technology development---especially at large aggregation scales of analysis 
\citep{Jaffe2000,Tussen2000,Verbeek2002,VanLooy2003,Harhoff2003,Roach2012}.\footnote{The inverse feedback, namely the influence of technology on science, 
has been proxied using patents cited by scientific publications~\citep{Hicks2000,Glanzel2003}, which have however a less clear interpretation than references in the opposite direction \citep{BarIlan20081}.} 

Notably, various studies \citep{Narin1997,Meyer2000,Meyer2000b,Callaert2014} conclude that interactions between science and technology 
are much more complex (and reciprocal) than a linear model of knowledge flow would suggest. Indeed, scientific and technological activities mutually benefit from such interactions: 
patent-cited papers perform better in terms of standard bibliometric indicators \citep{Meyer2010}, 
and patents that cite journal research articles receive more citations---possibly because their influence diffuses faster in time and space \citep{Sorenson2004}.
Recently, \citet{Ahmadpoor2017} corroborated these observations using the network of references listed in papers and patents, 
which also allows to quantify the descriptors around basic and applied scientific research.
Besides giving insights on specific knowledge creation patterns, citation-based indicators can also offer a broader and more systematic view on science-technology relations, 
potentially addressing policy relevant issues on how to efficiently shape national innovation systems. 
Indeed, when performed at the level of nations, {\em science intensity} has been often compared to technological productivity (\ie, number of patents per capita), 
finding a positive relation in specific technological fields (biotechnology, pharmaceuticals, organic fine chemistry and semiconductors) \citep{Verbeek2003,VanLooy2003,VanLooy2007}. 
In particular, {\em science intensity} appears to be relevant for scientific sectors having a sufficient body of knowledge~\citep{Tamada2006}.

As outlined above, in this work we add to the current discussion with the research aim of measuring and comparing the influence 
that the publication output of national scientific systems has on the global scientific and technological knowledge development. 
To task, we use two refined bibliometric indicators, based on citation that research papers accrued either from other papers or from patents. 
The first indicator, which we introduced in a previous work \citep{Cimini2016} in line with standard bibliometric impact metrics, 
is the average number of papers citations received by research articles produced by a country in a given time interval, normalized by the world average of such a quantity. 
This index is meant to measures the influence that the scientific body of knowledge produced by that country has onto subsequent development in science at the global level, 
and accordingly we will refer to it as {\em science relevance}. We then introduce a second indicator, {\em technology relevance},\footnote{The term ``technology relevance'' 
was originally introduced by \citet{Tussen2000} to study the contribution of scientific knowledge on technological development, yet without explicitly formulating an indicator.}
namely the average number of patent citations received by research articles produced by a country in a given time interval, again normalized by the world average of such quantity. 
As such, it is intended to measure the influence that the scientific activity carried out by that country has on technological developments worldwide. 
Once the science and technology relevance metrics are defined, we assess their temporal dynamics for several nations, and cross-compare these indicators 
to characterize national scientific systems features. We further relate the relevance metrics to national expenditures in R\&D. 
Overall, our work represents a step forward towards a quantitative characterization of the complex interconnection between science and technology in the knowledge creation and innovation process.

We remark that, though relying on the same source of information (citations to scientific publications contained in patents), 
our {\em technology relevance} index and the {\em science intensity} parameter proposed by \citep{Narin1997,Narin2000} are different: 
they just measure different aspects of knowledge transfer process. {\em Science intensity} is defined as the average number of scientific references listed in the patents published by a given country, 
and as such it measures how much the patent portfolio of that country is inspired by science. Thus, in this case the focus is on countries technological output and its link with worldwide science. 
{\em Technology relevance} instead measures how much the scientific knowledge produced by a country inspires the worldwide patenting activity: 
now the focus is countries scientific output and its link with worldwide technology. 
In this perspective, {\em technology relevance} can be seen as reflecting the knowledge outflow from proprietary science to technology, 
whereas, {\em science intensity} takes the reverse viewpoint by reflecting the knowledge inflow from science to proprietary technology.\footnote{Metrics relying on papers citations to patents 
\citep{Hicks2000,Glanzel2003} reflect the opposite flow from technology to science, hence measuring yet another aspect of the knowledge transfer process.}
Consistently, {\em science relevance} \citep{Cimini2016} measures how much the scientific output of a country inspires the worldwide scientific activity, 
thus considering a different knowledge flow---namely that internal to science.

\section{Materials and methods}\label{sec:Materials}

\subsection*{Data and basic statistics}

Basic statistics on scientific productivity and impact are collected from the SciVal platform (\url{www.scival.com}), a new API filtering data from Scopus (\url{www.scopus.com}) 
which allows downloading a variety of metrics aggregated at the level of nation, scientific sector and year. 
Note that the platform does not allow to obtain information on single documents, and SciVal policies prevent from downloading the whole database of papers. 
The collected data cover years from $1996$ to $2014$, and refer to $N_{c}=45$ nations and to $N_{d}=27$ scientific macro-sectors (according to the Scopus classification). 

The scientific production of a nation indicates the scholarly output authored by all the researchers affiliated with an institution of that nation.
Note that Scopus statistics are built using a full counting method.\footnote{In principle, papers can be assigned to nations using either a full counting 
or a fractional counting method~\citep{Waltman2016}. In the former, a publication co-authored by various nations is fully assigned to each of them, whereas, 
in the latter the assignment is weighted, \eg, by the fraction of authors or affiliations belonging to that nation (although no shared way to define weights exist).} 
While we are bound to use this approach, which is commonly adopted in the literature, recent contributions~\citep{Aksnes2012,Waltman2015} point to fractional counting 
as a more correct approach for country-level analyses. This is because larger countries tend to have a lower degree of international co-authorship among their publication output with respect to small countries. 
Hence we can expect that, in the following analysis, small nations with high level of internationalization will be to some extent favored to the detriment of large standalone countries.
Note also that Scopus (as other bibliometric databases) basically has a full coverage of English-written documents published in international peer-reviewed journals. 
Documents written in other languages and published in national journals are however important especially in Social Science and Humanities~\citep{Nederhof2006,Sivertsen2012}, 
which were thus excluded from our analysis (also as they are not particularly relevant for technological output), resulting in $N_d=22$ scientific sectors considered.
Whether language bias remains in the data after this selection is discussed below in section Results.
Finally, note that SciVal statistics are built without considering the journal where a paper is published, so that all citations bear the same value. 
While this approach may appear flawed at a first glance, we remark that taking the publication venues into consideration would convey all the biases 
coming from the exogenous and endogenous factors that enter in the effective publication mechanism, and that can follow different criteria than the real quality of the scientific work \citep{Waltman2016}.

Concerning patent data, SciVal provides aggregate statistics of patent citations to scientific literature.\footnote{Patent citations are usually classified as 
{\em Patent Literature} (PL), \ie, citations to other patents, and {\em Non Patent Literature} (NPL), namely citations to other kind of documents \citep{Callaert2006}. 
SciVal considers only NPL citations to journal research papers (amounting to around 60\% of total NPL).} Such data 
covers five patent offices: the World Intellectual Property Organization (WIPO), the Intellectual Property Owners association (IPO), 
the European Patent Office (EPO), the United States Patent and Trademark Office (USPTO) and the Japan Patent Office (JPO).
Note that the database lacks relevant offices such as the China Trademark and Patent Office (CTPO), which can lead to bias as patent applicants 
usually apply first at the home country office (and successively to other offices when deemed necessary)~\citep{Martinez2011}. 
Moreover, there are strong differences from office to office on regulation and practices of patent handling. 
For instance, JPO usually splits applications in several narrower patents~\citep{deRassenfosse2013}, while USPTO does not publish all the applications 
but enforces by law a patent applicant to refer to any prior documents known to him (the so called ``duty of disclosure'')~\citep{Verbeek2002}. 
This results in different citation frequencies among USPTO, EPO and JPO patents (see Fig. \ref{fig:pat_off}). 
Note also that SciVal does not allow to group patents accordingly to families, hence there could be repetitions in the data if the same invention is patented in different offices. 
While applying for a patent at multiple offices could be seen as a measure of success, it can also be the source of potential bias. 
Overall, however, the granularity of the data (\ie, the multitude of patent references we consider) and the aggregation of various offices 
are likely to average out eventual distortions, thus mitigating all of these problems (as demonstrated by the robustness analysis reported in the Supplementary Materials section).

Keeping all the described issues of the dataset in mind, we collect from SciVal the aggregate statistics on citations that research papers from journals, \ie, scientific documents, 
receive from other papers as well as from patents. Thus, for each nation $i$, scientific sector $\alpha$ and year $t$, 
we consider the whole corpus of the journal research papers produced, amounting to $[DOC]_{i\alpha}(t)$, and record the following basic metrics: 
\begin{itemize}
 \item $[CIT]_{i\alpha}(t)$, the number of citations these papers receive from other research papers, \ie, the number of times these papers appear as references listed in other research papers; 
 \item $[PCC]_{i\alpha}(t)$, the number of citations these papers receive from patents ({\em patent-citation count}), \ie, the number of times these papers appear as references listed in patents; 
 \item $[PCD]_{i\alpha}(t)$, the size of the subset of these papers that are cited by patents ({\em patent-cited documents}), \ie, how many of these papers are listed as references in patents. 
\end{itemize}
Overall, the data we collected refer to $33\,215\,296$ journal research papers, $656\,378\,663$ paper citations, $7\,949\,465$ patent citations, and $1\,487\,612$ patent-cited research papers.

We complement this information with measurements of national expenditures in research and development (R\&D), 
collected from the Organization for Economic Cooperation and Development (OECD, \url{www.oecd.org}). 
Data refer to GERD (Gross Expenditures on R\&D) values for $N_{f}=44$ nations from $1981$ to $2015$,\footnote{All expenditures 
are expressed in terms of current purchasing power parity (in millions of US dollars).} divided into three subcomponents depending on the funded sector: 
BERD (Business Expenditure on R\&D) for the business sector, HERD (Higher Education Expenditure on R\&D) for basic research performed in the higher education sector, 
and GOVERD (Government Intramural Expenditure on R\&D) for the government sector (we remand to \citet{OECD2002} for the official definition of these quantities). 
In the following, we consider HERD and BERD only, excluding GOVERD as it concerns the government research sector---which is often mission-oriented 
and therefore less related to scientific productivity, be it patent-cited or not~\citep{OECD2002,Leydesdorff2009}. 
Note that data coverage is not uniform, with several missing values before $1995$ and from 2009 onwards. 
Additionally, HERD is available only for $37$ nations while BERD only for $34$ nations. 
We therefore restrict the analysis on R\&D expenditures to years $1996-2008$ (compatibly with the SciVal database and data availability), 
and to the $N_f=34$ nations whose data is available~\footnote{To compensate for the few missing values in the restricted database, we use linear interpolation on the available data.}.

\subsection*{Relevance indicators}

As stated in the Introduction, our aim in this paper is to measure the influence that the scientific body of knowledge produced by a nation in a given scientific sector 
has onto subsequent development both in science and technology, at the global level. We build these indicators using the information provided by the citations that research papers 
produced by a nation (which we take as proxy of that nation's scientific system output) receive either from other papers or from patents.
Bearing in mind all the caveats mentioned in the Introduction, the underlying idea is that research papers cite to acknowledge existing work on which they build on, 
and patents to acknowledge contribution to the inventive process. Thus, we can use paper/patent references to the papers produced by a nation (\ie, at a large aggregation scale) 
to assess the relevance of the scientific findings by that nation for future developments in science/technology.

To measure the impact of the scientific production of a nation on subsequent global scientific activity, we use standard scientometrics tools 
based on shares of accrued scientific citations~\citep{Waltman2016}. In particular, we define the {\em science relevance} index as the {\em citation share over document share}, 
defined in~\citet{Cimini2016}, namely the average number of paper citations received by research articles produced by a country in a given year, normalized by the world average of such quantity:
\begin{equation}
	\textit{Sci } [Csh/Dsh]_{i}(t) = \left(\frac{\sum_{\alpha} [CIT]_{i\alpha}(t)}{\sum_{j\alpha}[CIT]_{j\alpha}(t) } \right) \Bigg/ \left(\frac{\sum_{\alpha} [DOC]_{i\alpha}(t)}{\sum_{j\alpha}[DOC]_{j\alpha}(t) } \right) 
	= \frac{1}{A_{sci}(t)}\frac{\sum_{\alpha} [CIT]_{i\alpha}(t)}{\sum_{\alpha}[DOC]_{i\alpha}(t)},
	\label{eq:SciSuccess}
\end{equation}
where the average paper citations per document $A_{sci}(t)=(\sum_{j\alpha} [CIT]_{j\alpha}(t)/(\sum_{j\alpha}[DOC]_{j\alpha}(t))$ allows for proper time normalization 
(papers published more recently had less time to attract citations~\citep{Medo2011}). 
Note that in the above formula all papers are given the same weight, whereas, other metrics use a field normalization approach by giving different weight to papers 
belonging to different scientific sectors~\citep{Waltman2011}. Remarkably, the different approaches found in literature lead to practically the same results 
when applied at the aggregate level of nations~\citep{Cimini2016}. 

To measure the influence of the scientific production of a nation on the global technological development, we adopt the same reasoning used for eq.~(\ref{eq:SciSuccess}). 
We thus define the {\em technology relevance} index by replacing, in the above expression, citations from scientific papers with citations from patents. 
In this way, we obtain the average number of patent citations received by research articles produced by a country in a given year, normalized by the world average of such quantity:
\begin{equation}
	\textit{Tech } [Csh/Dsh]_{i}(t) = \left(\frac{\sum_{\alpha} [PCC]_{i\alpha}(t)}{\sum_{j\alpha}[PCC]_{j,\alpha}(t) } \right) \Bigg/ \left(\frac{\sum_{\alpha} [DOC]_{i\alpha}(t)}{\sum_{j\alpha}[DOC]_{j\alpha}(t) } \right) 
	= \frac{1}{A_{tech}(t)}\frac{\sum_{\alpha} [PCC]_{i\alpha}(t)}{\sum_{\alpha}[DOC]_{i\alpha}(t)},
	\label{eq:TechSuccess}
\end{equation}
where again time normalization is achieved trough the average patent citations per document $A_{tech}(t)=(\sum_{j\alpha} [PCC]_{j\alpha}(t))/(\sum_{j\alpha}[DOC]_{j\alpha}(t))$.
As for the case of scientific relevance, using a field normalized variant of the technology relevance index leads to very similar results (see Fig. \ref{fig:MNCsh}). 
Notably, this happens despite patent citations can be concentrated on selected scientific sectors.

Note that both relevance indicators reflect the influence of the publications produced by a country, taken as a proxy of its scientific systems output, 
and not the scientific or technological relevance of the country per se (which many depend on many other factors).
We also remark that the proposed metrics are built to be independent from country size (\ie, they are intensive quantities), with the underlying rationale that whenever a nation receives 
a larger share of citations compared to the fraction of papers it publishes, it is producing science that has a greater impact than the world average. 
As compared to the average-based indices already proposed in the literature~\citep{Waltman2016}\footnote{An alternative approach to average-based indicator 
is represented by percentile-based indicators~\citep{Waltman2013}, which are less sensitive to outliers given by highly cited publications~\citep{Aksnes2004}. 
Yet, when performing analyses at large scales (\eg, for nations, wide scientific areas, and long time windows), the law of large numbers acts by smoothing out the distortions 
due to such outliers~\citep{Cimini2016}.}, the advantages of the specific formulation we adopt here are the minimization of fluctuations due to small scientific sectors, 
and the independence on the classification scheme used for science, which we pay by loosing a proper field normalization. 

We conclude the section with a discussion on the statistical robustness of the proposed indicators. 
Eqs.~(\ref{eq:SciSuccess}) and~(\ref{eq:TechSuccess}) are computed with numbers of very different magnitude, 
as the number of scientific references research articles obtain from papers is far greater than the number of references from patents. 
This may in principle cause large fluctuations for the technology relevance index. 
Nevertheless, the level of aggregation we use (citations received by all research papers of a country by worldwide patents in a given year) 
is large enough for the law of large numbers to hold, and so to obtain reliable statistics. This is demonstrated in the section below by the smooth trends of the indicators 
over the considered time span. Note also that the normalization we use for both indicators make them comparable, whatever the magnitude of the original data.

\section{Results and Discussion}\label{sec:Results}


\begin{figure}[h!]
	\centering
	\includegraphics[scale=0.5]{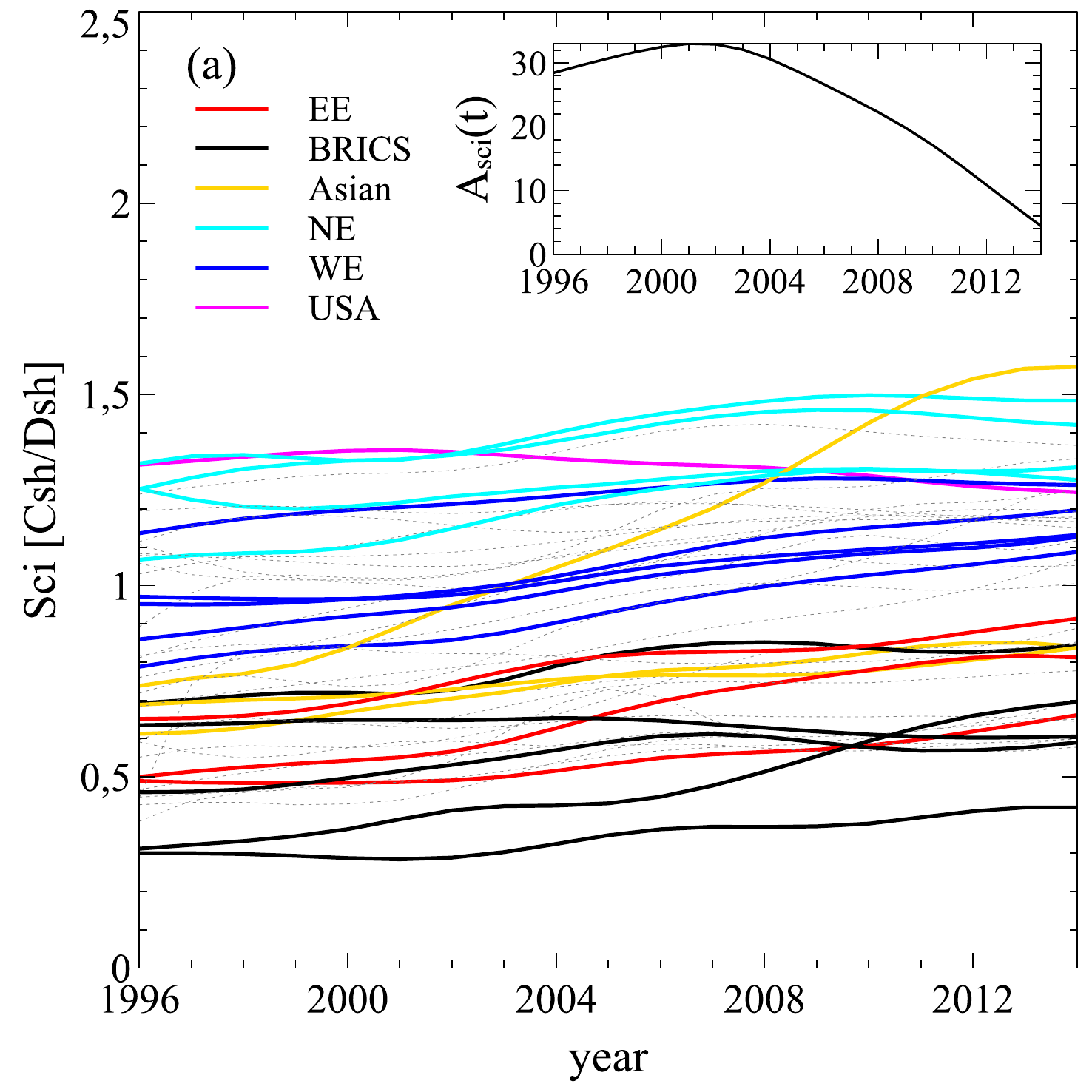}
	\includegraphics[scale=0.5]{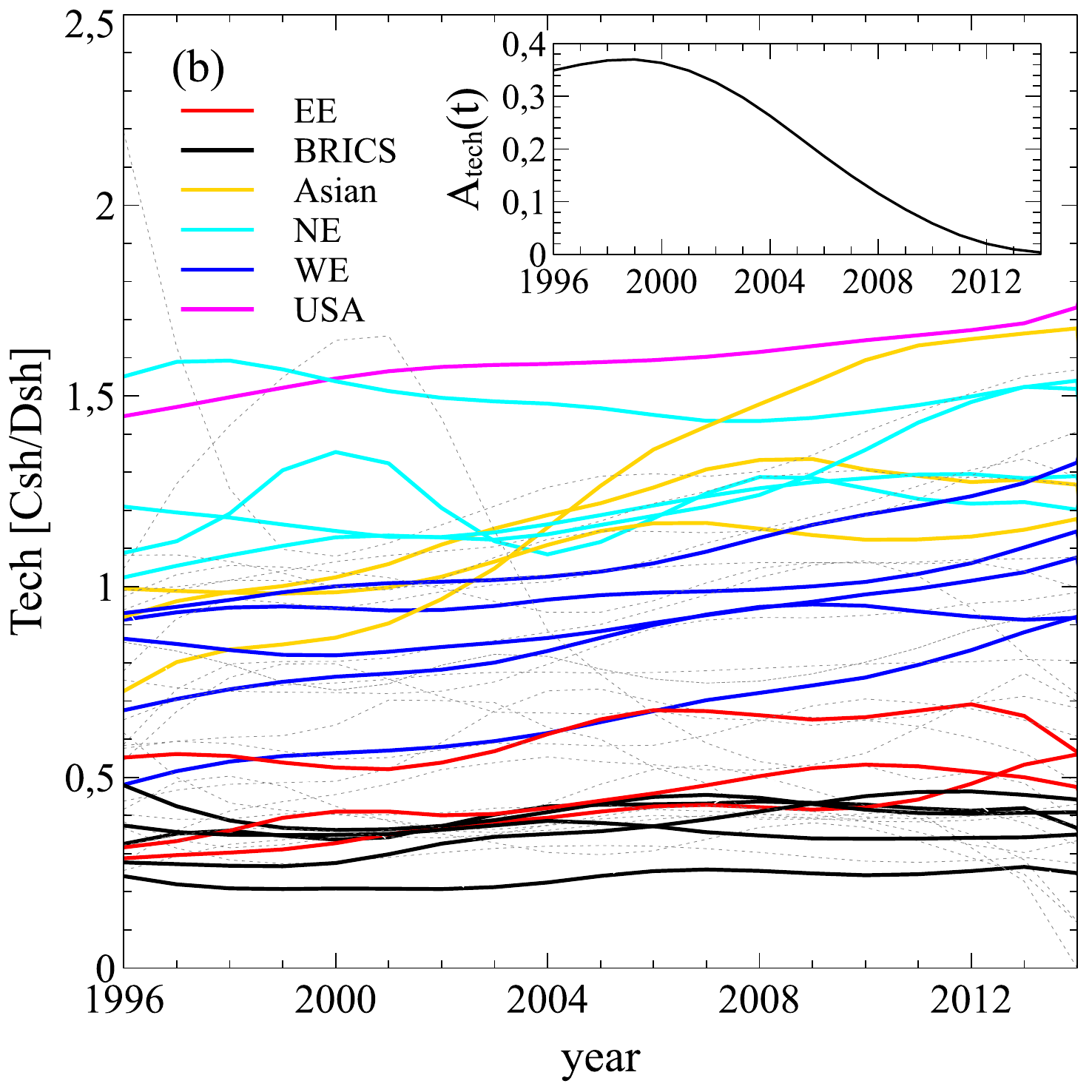}
	\caption{Relevance indices of different nations over years $1996-2014$. 
			Panel (a): {\em science relevance} defined by eq. \eqref{eq:SciSuccess}. Inset: time normalization factor $A_{sci}(t)$.
			Panel (b): {\em technology relevance} defined by eq. \eqref{eq:TechSuccess}. Inset: time normalization factor $A_{tech}(t)$.}
	\label{fig:TemporalSuccesses}
\end{figure}

The science and technology relevance metrics of nations, respectively obtained with equations~\eqref{eq:SciSuccess} and~\eqref{eq:TechSuccess}, are shown in Figure~\ref{fig:TemporalSuccesses}.
Line colors correspond to different cultural, economic and geographical regions, for which we report representative countries (other nations are reported in gray): 
magenta identifies the United States, blue denotes Western Europe (France, Germany, United Kingdom, Italy, Spain), black are BRICS countries (Brazil, Russia, India, China, South Africa), 
cyan denotes Northern Europe (Belgium, Netherlands, Sweden, Switzerland), red denotes Eastern Europe (Czech Republic, Hungary, Poland) and yellow identifies Asian countries 
(Japan, South Korea, Singapore, Taiwan). Concerning {\em technology relevance}, shown in panel $(b)$, a clear separation emerges between very efficient nations 
(\ie, USA, Switzerland and Singapore in the late years), Europe and developed Asian countries, and the rest of the world. This pattern is observed also when field normalization is employed 
(Fig. \ref{fig:MNCsh}), and when the analysis is restricted to individual patent office data (to control for home advantage effects, Figs. \ref{fig:no_home_adv} and \ref{fig:pat_off}) 
as well as to specific scientific sectors (Fig. \ref{fig:sectors}). The trend of the time normalization coefficient, \ie, the average number of patent citations per document over all countries 
(shown in the inset of panel $(b)$) indicates a characteristic time scale for patents citations of about $10-15$ years, thus longer than that for scientific citations (as shown in the inset of panel $(a)$). 
Indeed, papers need time to attract citations from other papers, and even more time to get citations from patents. 
One of the reasons is that patent applications are not processed in real time but with a delay of $30-40$ months, depending on the patent office~\citep{Ackerman2011,Mejer2011}\footnote{This 
reduces the validity of the available data in the most recent years~\citep{Hall2001}}. Panel $(a)$ shows instead {\em science relevance} for the different nations. 
Although it is still possible to find a separation between geographical areas, no particular gap is observed between Europe and the most efficient nations.


\begin{figure}[h!]
	\centering
	\includegraphics[scale=0.5]{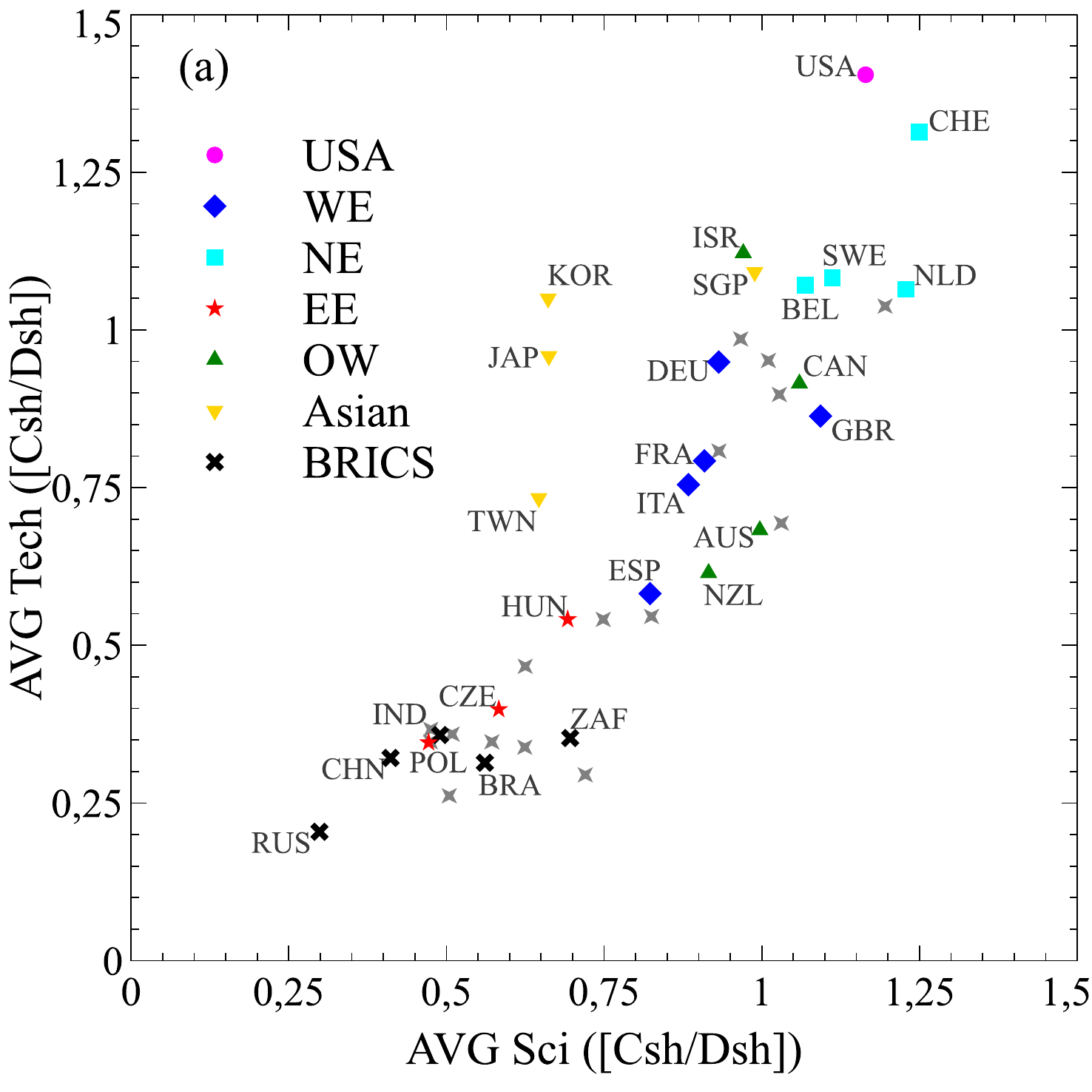}
	\includegraphics[scale=0.5]{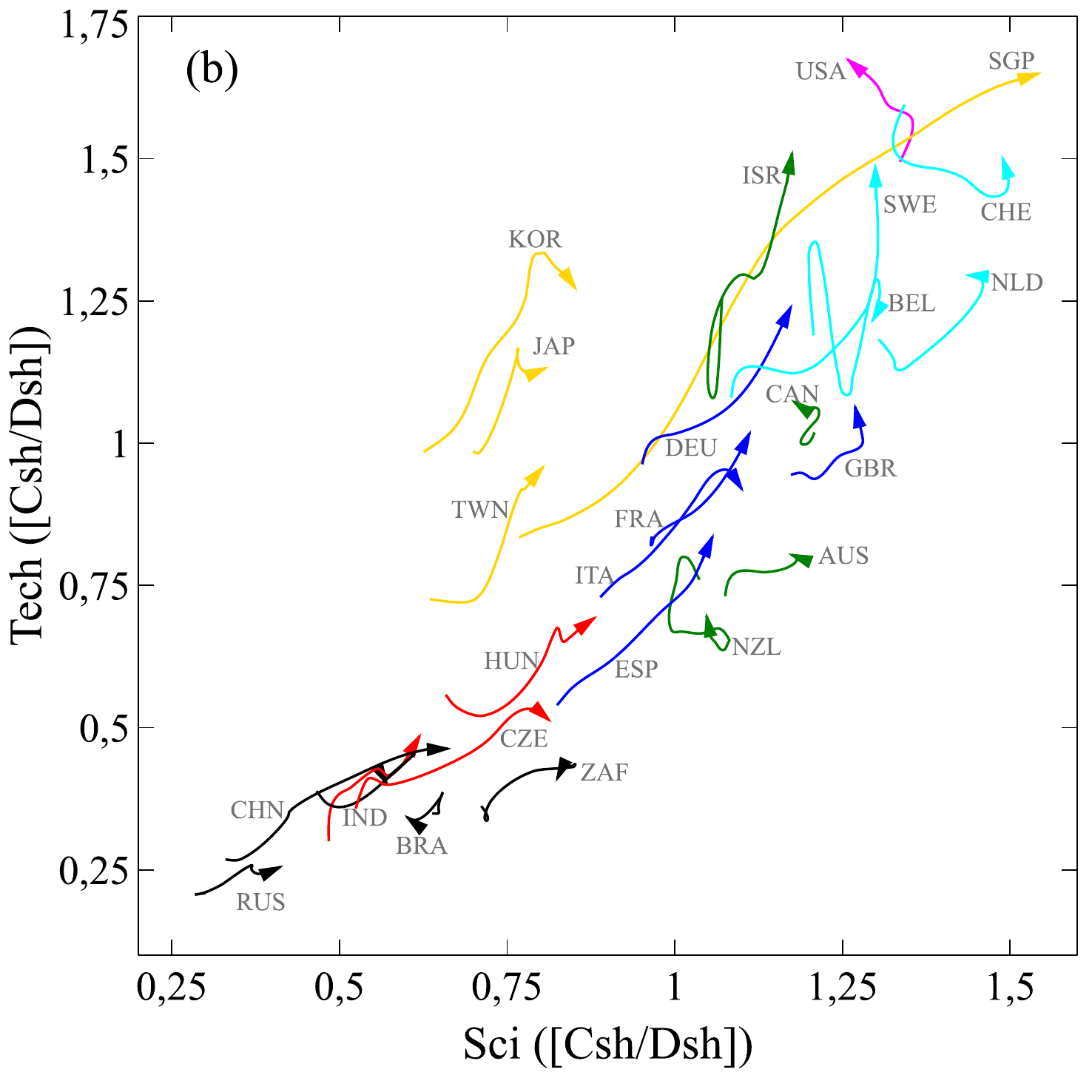}
	\caption{{\em Science relevance} versus {\em technology relevance} of nations over years $1998-2012$: average values 
	(panel (a)) and temporal trajectories with arrows indicating the time flow (panel $(b)$). 
	Besides the regions described for Fig~\ref{fig:TemporalSuccesses}, here we highlight also other western countries (Australia, Canada, Israel, New Zealand) in green. 
	Temporal trajectories are realized with a data smoothing procedure using a five year window; as such, we do not show values for the first and last two years of the data.}
	\label{fig:AVG_Trajectories_CshDsh}
\end{figure}

To better understand the relation between science and technology relevance metrics, Figure~\ref{fig:AVG_Trajectories_CshDsh} shows the scatter plot of these values for the various nations. 
Panel $(a)$ reports values averaged over years $1998-2012$. Indeed, we see that the two metrics are not independent and, as expected, are highly correlated.
Remarkably, the plot highlights a cultural and geographical separation. 
Developing nations (such as BRICS countries) are located in the bottom left section of low relevance both in science and technology. 
Then the central region is populated, in sequence, by Easter European countries and Western European countries, ending with North Europe and the top performers Switzerland and USA. 
Asian countries lie slightly off the diagonal, featuring a higher technological influence than scientific impact. 
This discrepancy may be induced from the aforementioned JPO practice of splitting patents into more applications, 
resulting in more patent citations for Asiatic countries relying on this office as compared to other nations with similar scientific relevance values. 
Note also the position of China, which has low value of scientific relevance because of the intensive nature of the index 
(China still has a low citation ratio in science) as well as a low technological relevance value also due to the lack of the CTPO in our dataset. 
We remark that metrics intensivity is the main reason for small countries like Switzerland and Singapore outperforming other large nations like China and Russia. 
This feature is however meaningful when considering that the former countries have a rather efficient and more applied research system. 
The latter country indeed have a much larger impact overall, which is captured by extensive metrics like total number of documents or citations.

Moving further, panel $(b)$ shows the trajectories of the different nations in the plane of {\em science relevance} and {\em technology relevance}. 
To understand the plot, we first notice that both the relevance measures are based on citation shares over document shares, 
and since the world's shares are necessarily equal to one, the relevance indices cannot grow for all nations at the same time. 
Most of the nations, especially those in the center of the diagram, show a positive trend both for scientific relevance values, compensated by the notable exception of the USA, 
and for technological relevance values, compensated instead by the decrease of some under-developed countries (not shown in the plot), as well as by that of Japan and Korea 
at the end of the time span considered. We then see that developing nations do move towards regions of higher relevance, but in a chaotic manner (except for China that moves smoothly). 
On the contrary, the motion of Western countries towards the region of highest scientific and technological relevance values is more laminar (and an extraordinary improvement 
is observed for Singapore). Notably, this heterogeneous dynamics reflect those found in the study of economic development~\citep{Cristelli2013,Cristelli2015}. 

Figure~\ref{fig:AVG_Trajectories_CshDsh} also allows to dwell into eventual language biases affecting our data and metrics. 
We observe that anglophone countries like Australia, Canada, New Zealand and United Kingdom perform similarly to Western Europe countries of comparable size (France, Germany, Italy, Spain). 
Hence, were language biases at work, they would have an effect only on selected countries, putatively large nations like China and Russia. 
However we remark that language biases would substantially affect our indicators basically if, for a nation: 
i) the number of papers written in other languages is comparable with the number of papers written in English, and 
ii) the citations received by the former papers are more than those received by the latter. These conditions (especially the second one) is hardly realized in practice in any country.


\begin{figure}[h!]
	\centering
	\includegraphics[scale=0.5]{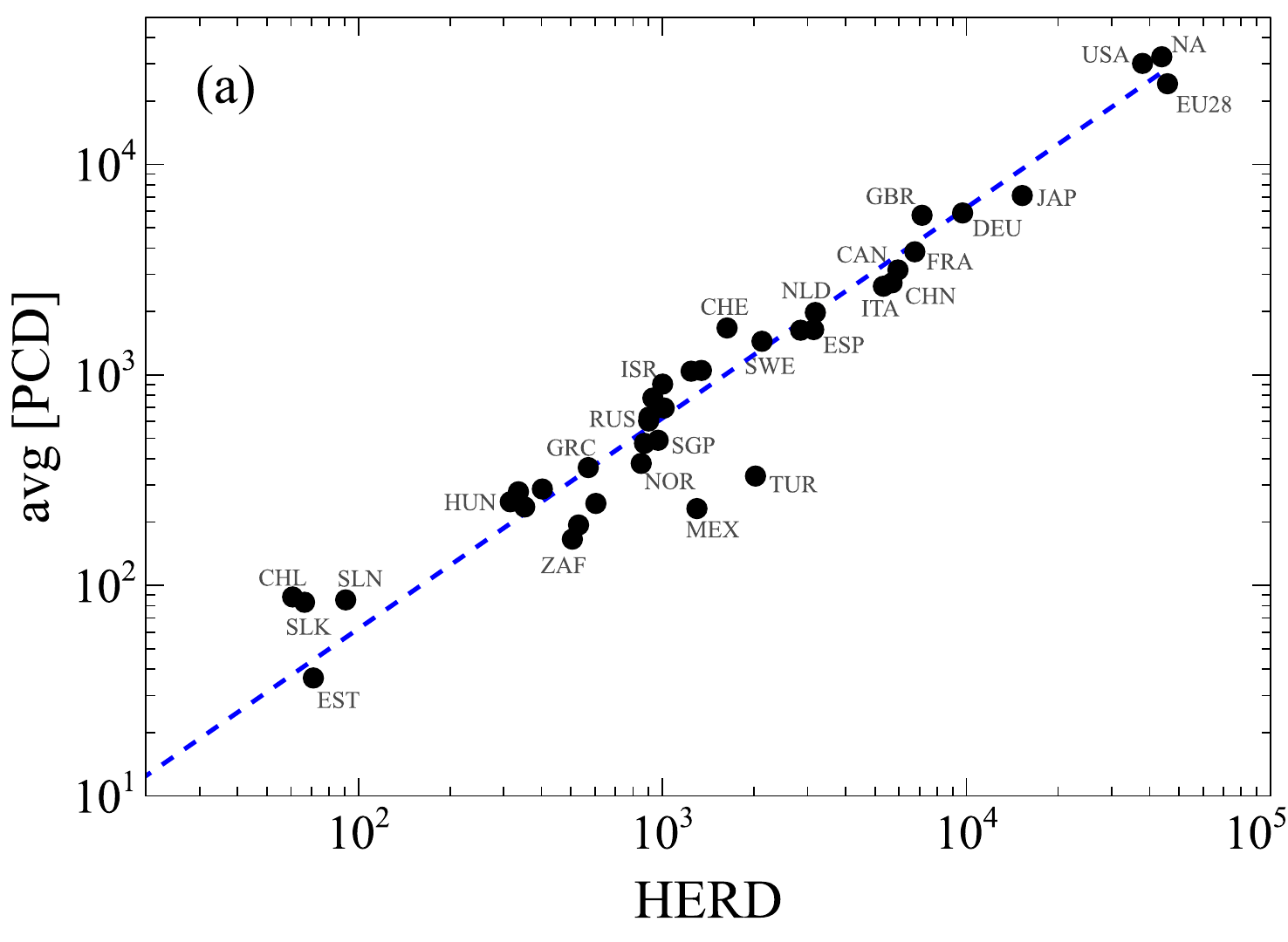}
	\includegraphics[scale=0.5]{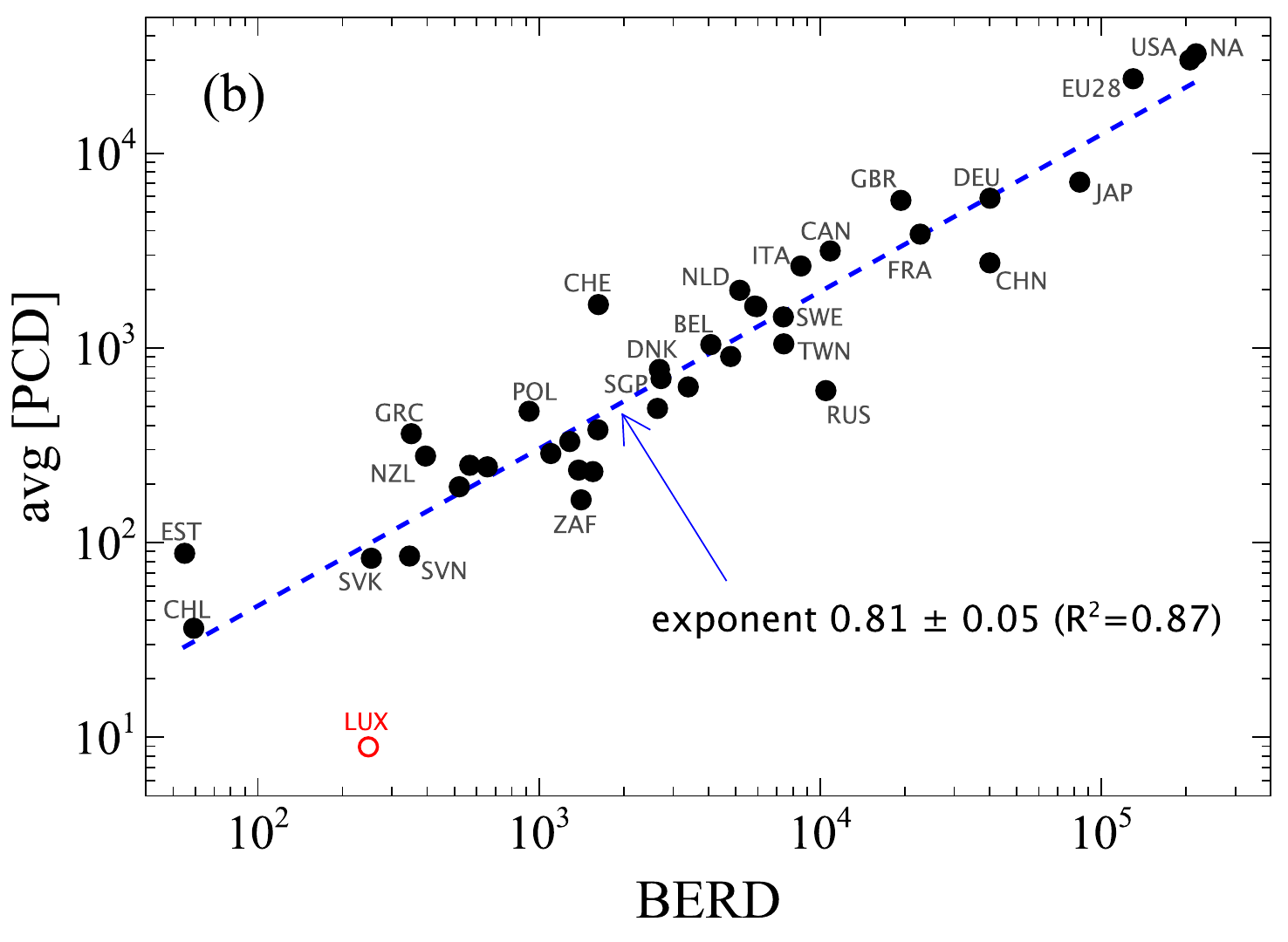}
	\caption{Scatter plot of HERD (panel $(a)$) and BERD (panel ($b$) R\&D funding versus the amount of patent-cited scholarly output, averaged over years $1996-2008$. 
	Points correspond to different nations, and the blue dashed lines mark the best regressions (linear with $R^2=0.84$ for HERD, power-law with exponent $0.81 \pm 0.05$ and $R^2=0.87$ for BERD).} 
	\label{fig:HERDandBERD_PCD}
\end{figure}

We conclude with an analysis of patent citation statistics with respect to national expenditures in R\&D. 
In particular, we consider HERD as usually done in studies focused on bibliometric scientific outputs, as well as BERD 
which is supposedly more related to patenting activity---and thus important for innovation and economic growth. 
Figure~\ref{fig:HERDandBERD_PCD} shows scatter plots of the patent-cited scholarly output versus these R\&D expenditures, both averaged from $1996$ to $2008$, for several nations. 
When HERD is considered (panel $(a)$), we observe a succession of points which is perfectly fitted with a straight line. 
This result is expected if we assume that the number of research articles cited by patents is an homogeneous subset of the scientific output of each nation, 
while overall scientific production scales linearly with HERD (see~\citet{Cimini2014} for what concerns scientific impact). 
More interestingly, when we consider BERD (panel $(b)$) and we exclude Luxembourg (the red outlier in the figure), the point are even more correlated with respect to HERD. 
However, the least square regression returns a power-law relation with exponent $0.81\pm 0.05$. 
This sublinear behavior suggests that the nations with the larger scientific production make more fundamental research 
and are thus at the boundary of knowledge---with only future applications and less possibility to induce immediate innovations through patenting activity.

\section{Conclusion}

In this work, using citation data from journal research papers as well as from patents, we have investigated the relevance of the scientific production of nations 
on science and technology at the global scale. Our approach of measuring the scientific capabilities of a country by looking at the technological sector beyond the scientific domain 
is in line with any recent theory of scientific production. To task, we designed a novel indicator for the influence of science on technology, in line with existing impact metrics for science, 
and show that a relation exists between scientific and technological relevances---which grow together for most of the nations considered in our study. 
This feature points to the positive feedback between knowledge production, discoveries and innovation. 

Geographical and cultural patterns emerge from the joint analysis of paper and patent citations: clusters of nations with similar scientific production systems do appear. 
For instance, we find that Northern European countries (including Switzerland) have a highly influential scientific system, 
along the observation made by \citet{Cimini2016}, and their scientific production has an even larger influence on patent literature.  
We also observe a gap between Western Europe and USA that does not emerge by looking at scientific citations alone, 
and which can be due to a more optimized and integrated National Innovation System in the USA. 

Moving futher, we find that the amount of national scientific production with technological relevance (\ie, cited by patents) 
strongly correlates with the total expenditure in R\&D by business institutions and enterprises. 
A possible explanation for the sub-linear behavior observed in this case may be that the larger such expenditures, 
the larger the research efforts in the private sector aimed at fundamental research with no immediate technological spin-offs.
Overall, the good agreements we find between input in resources for R\&D and relevance of knowledge output suggests that our indicators can be reliably used 
to make quantitative assessments on National Innovation Systems features.

We recall that there are two main reasons for small countries generally winning the competition with large countries in our study. 
The first one comes from data and is the use of full counting to assign citations from internationally co-authored papers to countries. 
Since larger countries tend to have a lower degree of international co-authorship among their publication output, they are to some extent penalized by this approach. 
The second reason comes from the definition of our relevance metrics, which are normalized to be intensive so that countries size effects are washed out. 
We use this approach precisely to measure not overall influence but rather efficiency and the application-oriented aspect of research systems.

Our study represents a preliminary and exploratory step in the understanding of the coupling and co-evolution of science and technology as different but interacting compartments of the innovation system. 
The research we presented in this paper can be extended in different directions. For instance, it could be interesting to carry out a similar analysis using a finer resolution level, 
that is, looking at scientific institutions instead of countries. Another possible direction would be to focus on citations from patents belonging to specific technological sectors, 
in order to study individual patterns of knowledge transfer at a lower aggregation scale. This would however require to assign technological codes to the paper-citing patents provided by SciVal 
by matching with them with those provided by other databases like PATSTAT. Alternatively, we could attempt to evaluate the interactions between individual scientific and technological sectors 
by looking at the specific co-occurrences of scientific and technological activities in each nation \citep{Pugliese2017}. In the long term, the challenge will be that 
of identifying the micro-determinants describing the complex interplay between scientific advancement, technological progress, economic development and societal changes 
within the multi-layered space of innovation and development.

\bigskip

This work was supported by the EU projects GROWTHCOM (FP7-ICT, grant n. 611272), CoeGSS (H2020-EU.1.4.1.3., n. 676547), and the Italian PNR project CRISIS-Lab. 
The funders had no role in study design, data collection and analysis, decision to publish, or preparation of the manuscript.
We thank the SciVal team for fruitful discussions and for providing access to data.


%

\newpage

\setcounter{table}{0}
\setcounter{figure}{0}
\setcounter{equation}{0}
\renewcommand{\thetable}{S\arabic{table}}
\renewcommand{\thefigure}{S\arabic{figure}}
\renewcommand{\theequation}{S\arabic{equation}}

\section*{Supplementary Materials}

\medskip

\begin{equation}
 \textit{Tech } [MNCsh]_{i}(t) = N_d^{-1}\sum_{\alpha}\left(\frac{[PCC]_{i\alpha}(t)}{\sum_j [PCC]_{j\alpha}(t)}\right) \Bigg/ \left(\frac{[DOC]_{i\alpha}(t)}{\sum_j [DOC]_{j\alpha}(t)}\right)
 \label{eq:TechSuccess2}
\end{equation}

\begin{figure}[h!]
	\centering
	\includegraphics[scale=0.5]{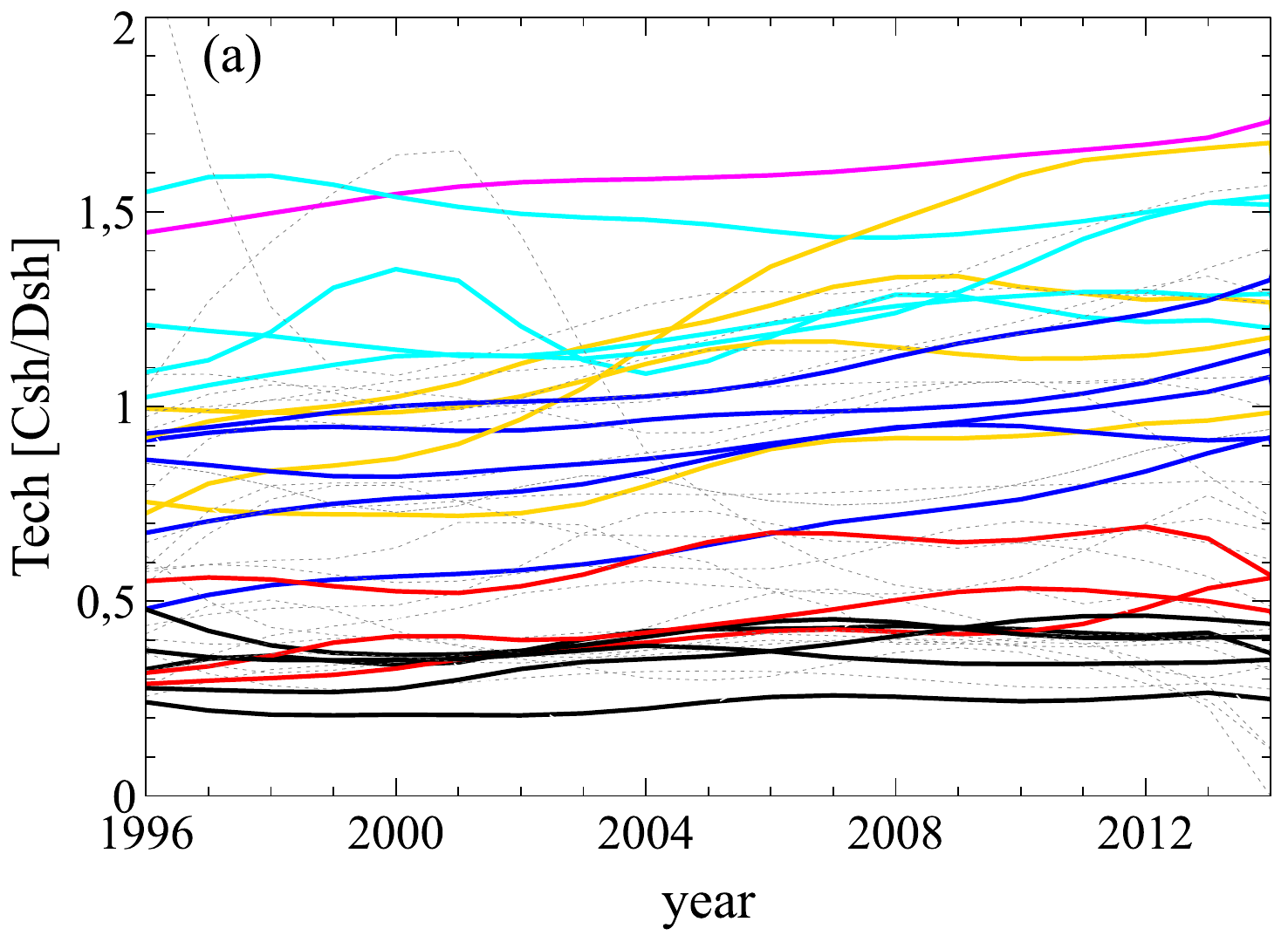}
	\includegraphics[scale=0.5]{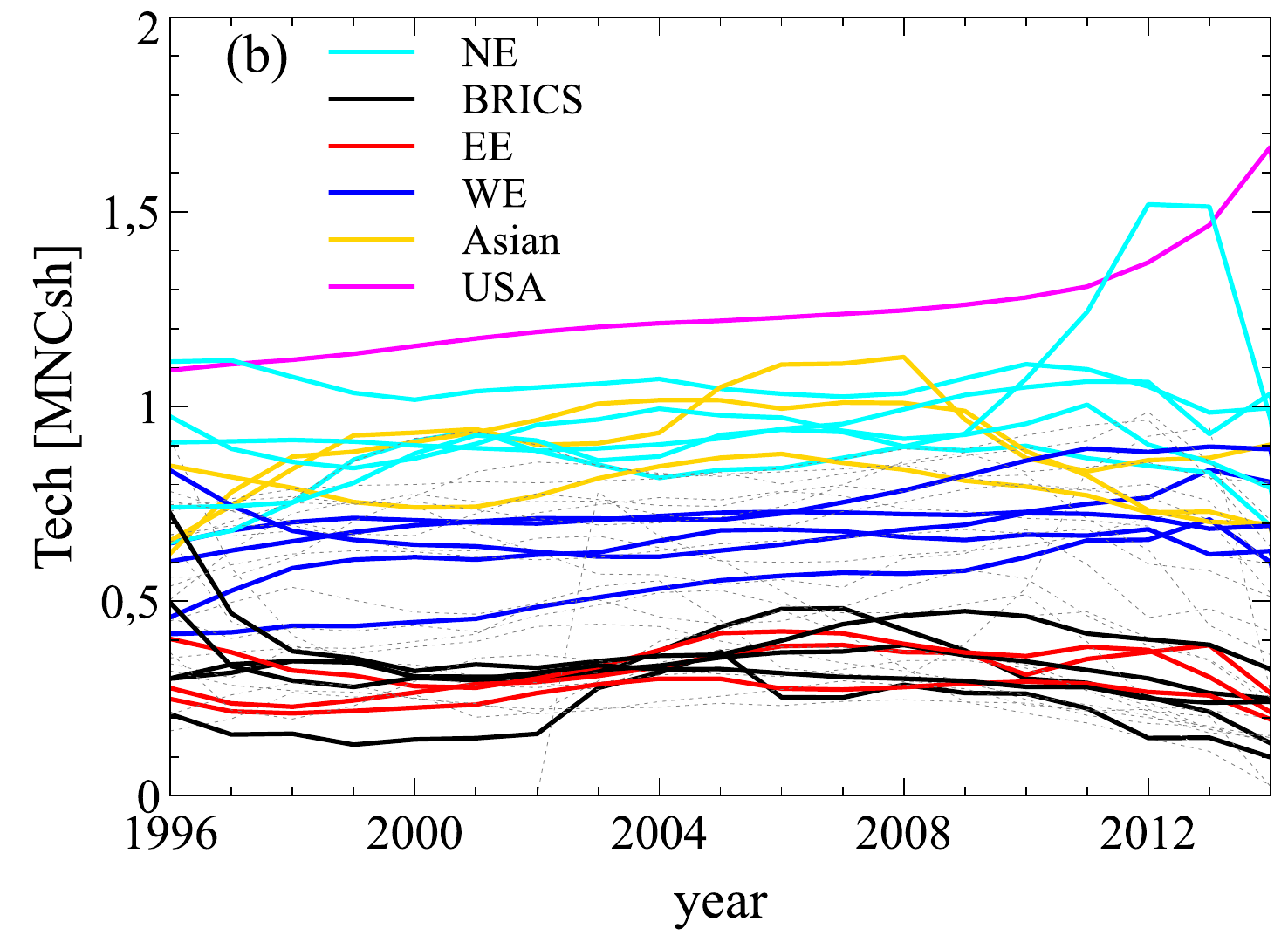}
	\caption{{\em Technology relevance} of nations, measured by the {\em citation share over document share} of eq.~(\ref{eq:TechSuccess}) of the main text (panel a), 
	and by the {\em mean normalized citation share} of eq.~(\ref{eq:TechSuccess2}) (panel b). 
	The latter index is field-normalized, and is analogous to the index defined in~\citet{Cimini2016} for scientific impact. 
	The overall Pearson correlation between the temporal averages of the two indices is very high, amounting to $0.98$ 
	(the Pearson correlation between the same metric variants for {\em science relevance} is $0.96$~\citep{Cimini2016}).}
	\label{fig:MNCsh}
\end{figure}

\begin{figure}[h!]
	\centering
	\includegraphics[scale=0.5]{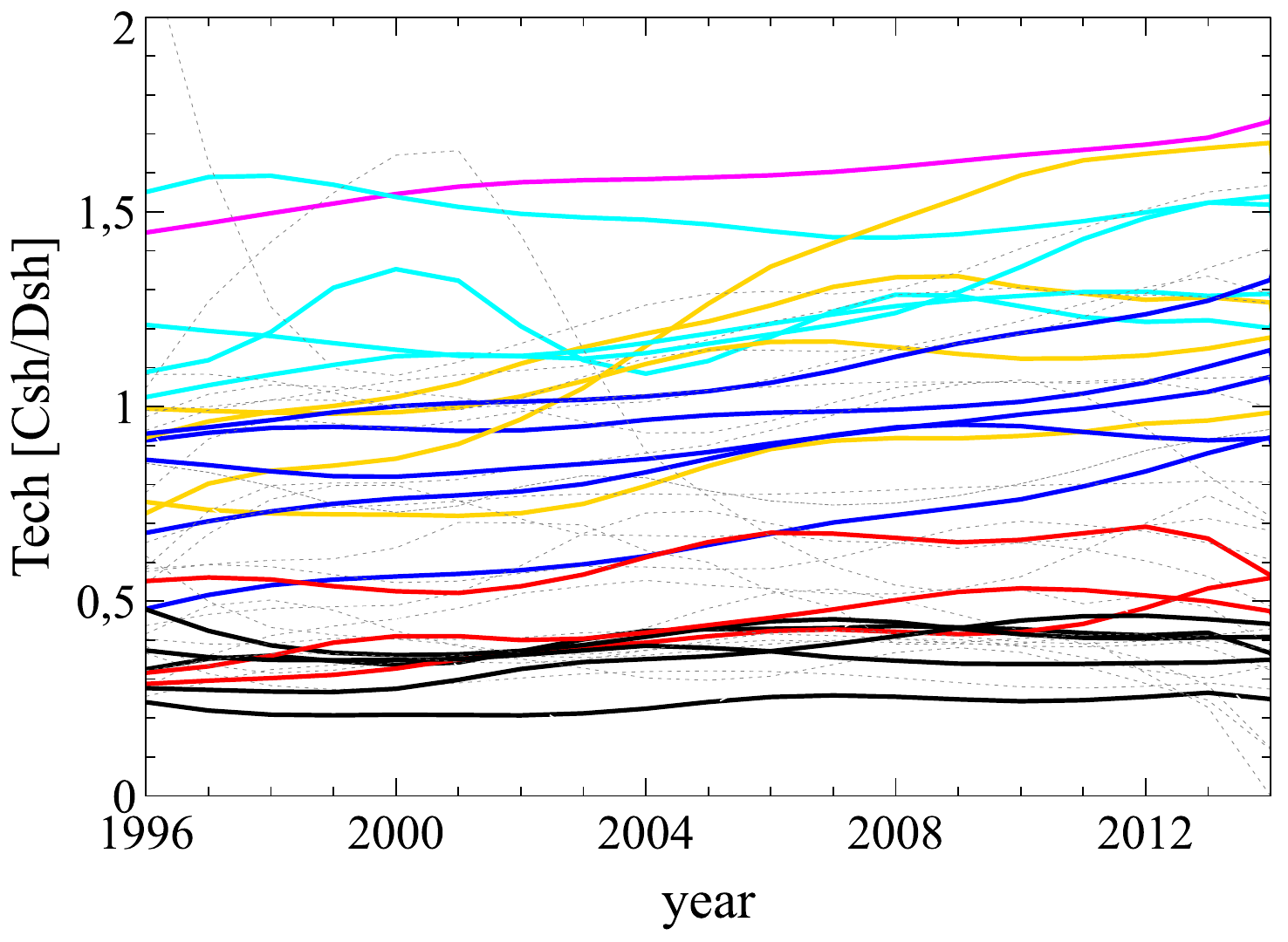}\\
	\includegraphics[scale=0.5]{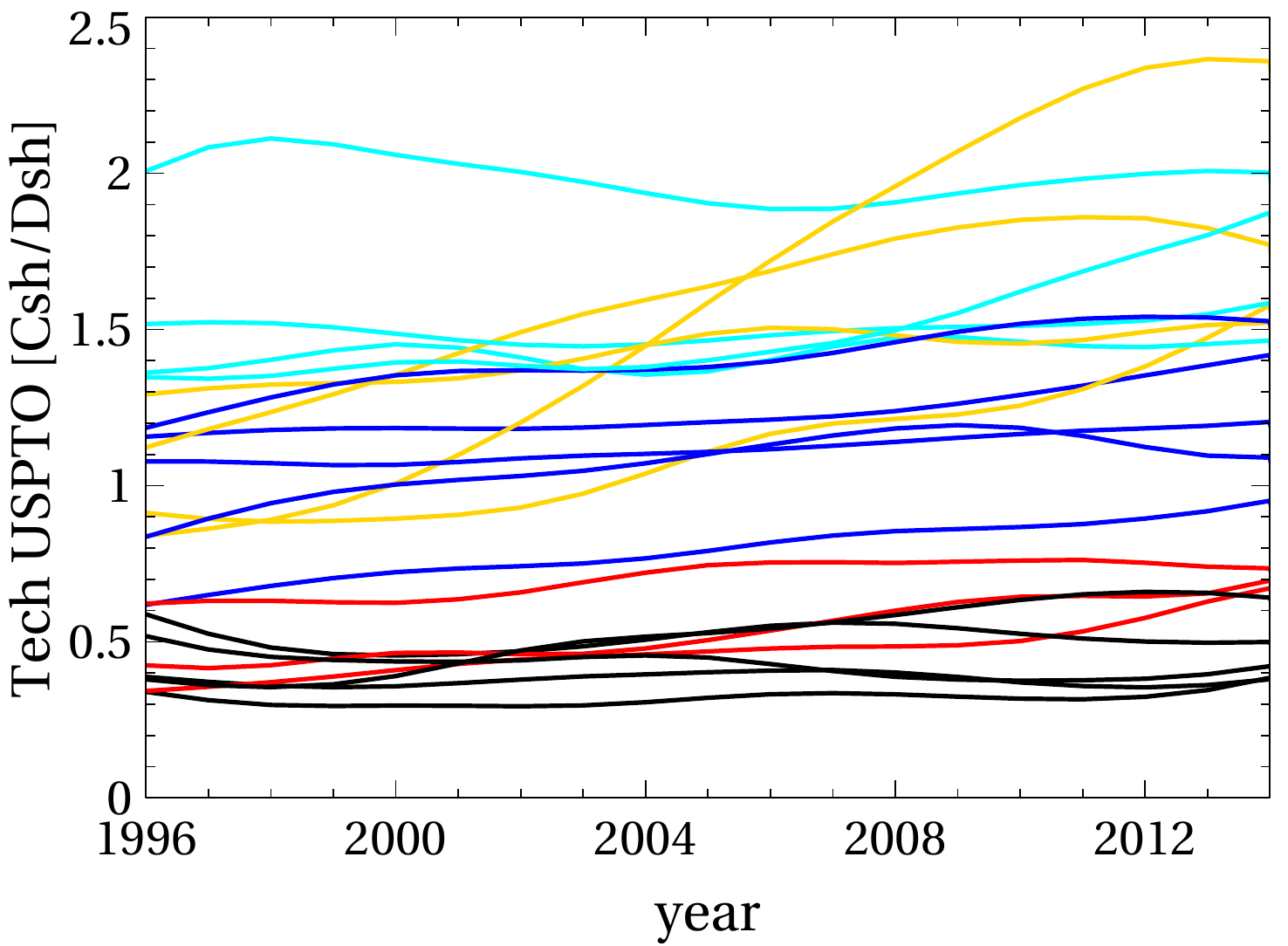}
	\includegraphics[scale=0.5]{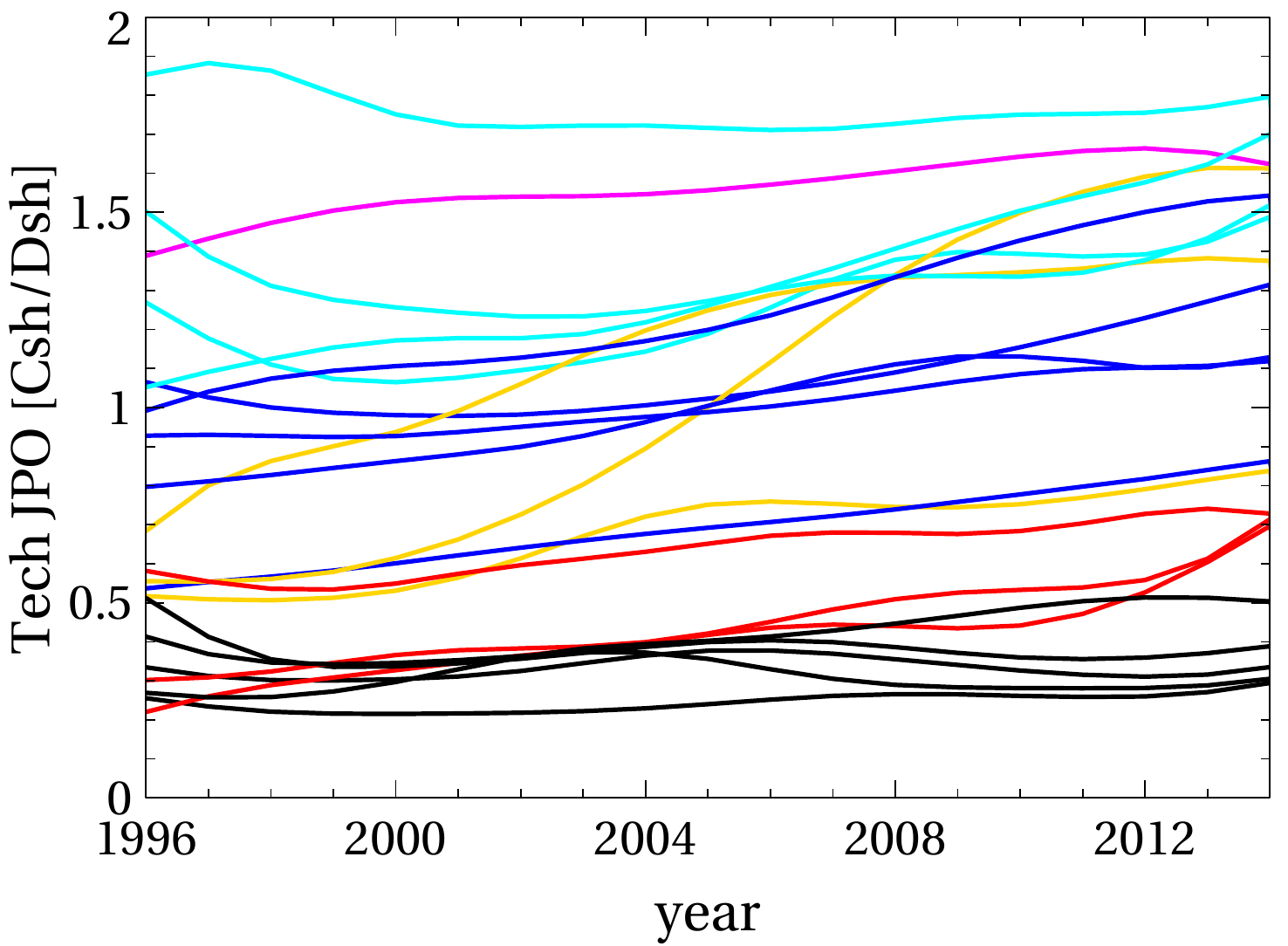}\\
	\includegraphics[scale=0.5]{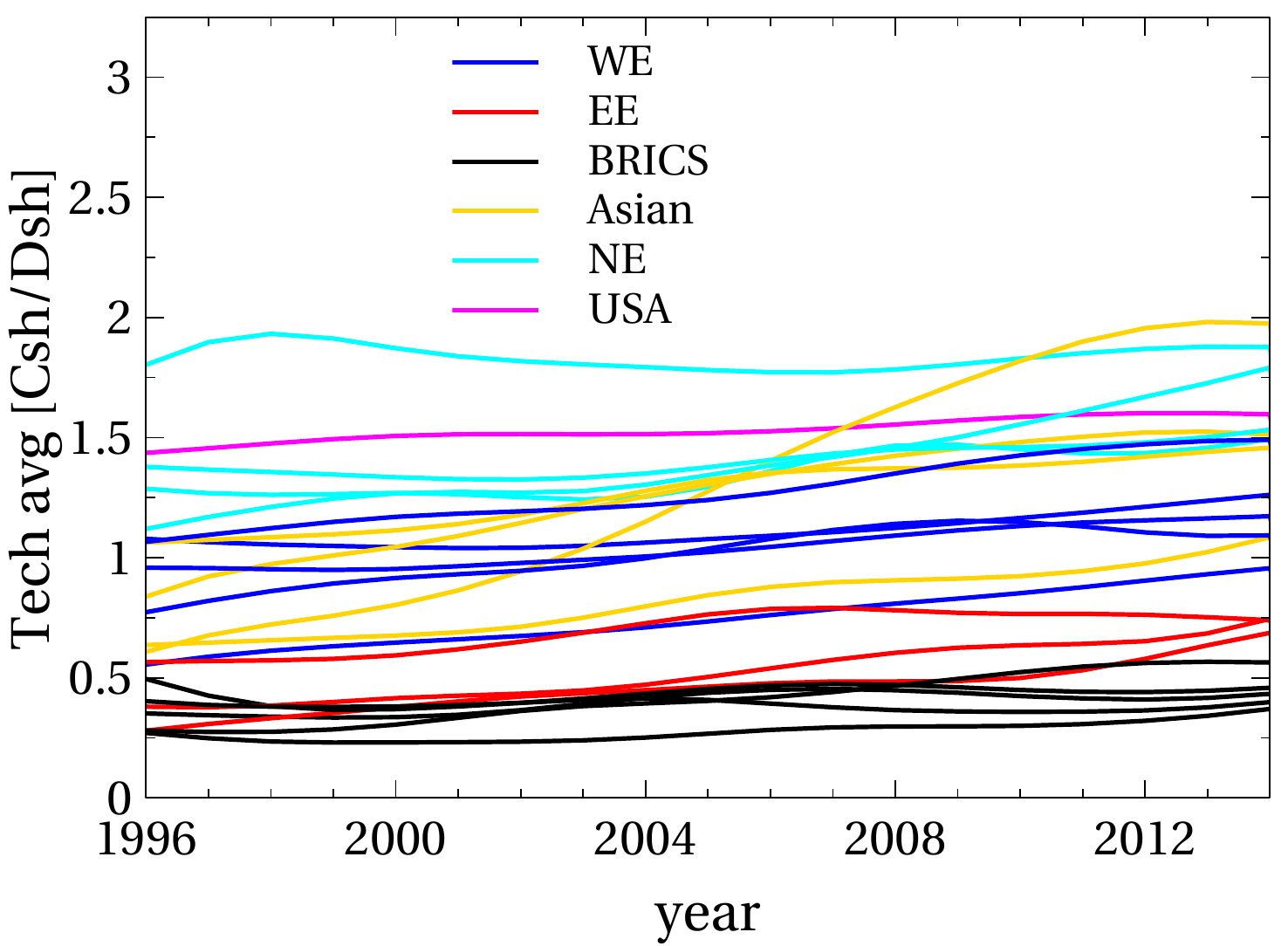}
	\caption{{\em Technology relevance} of nations computed using different patent data. In the top panel, all five patent offices are considered (it is the same of Fig. \ref{fig:TemporalSuccesses} 
	in the main text). In the central panels, only USPTO (left panel) and JPO (right panel) data are employed. In these two cases, the home country of the considered patent office 
	is not included in the analysis. These figures show that while home countries of patent offices do benefit from home advantage 
	(\eg, USA gets close to Europe when only JPO is considered), trends remain similar and the ranking of countries is robust. 
	For instance, Switzerland (and also USA) still occupies top positions, and Singapore enjoys a rapid ascend in relevance values (which is striking in the USPTO case).
	The bottom panel shows the technology relevance index computed as average on all patent offices but that of the home country (if present) and IPO. 
	For instance, in the case of USA we average over EPO, JPO and WIPO only, while for China or Russia we average over all four offices (\ie, including USPTO).}
	\label{fig:no_home_adv}
\end{figure}

\begin{figure}[h!]
	\centering
	\includegraphics[scale=0.5]{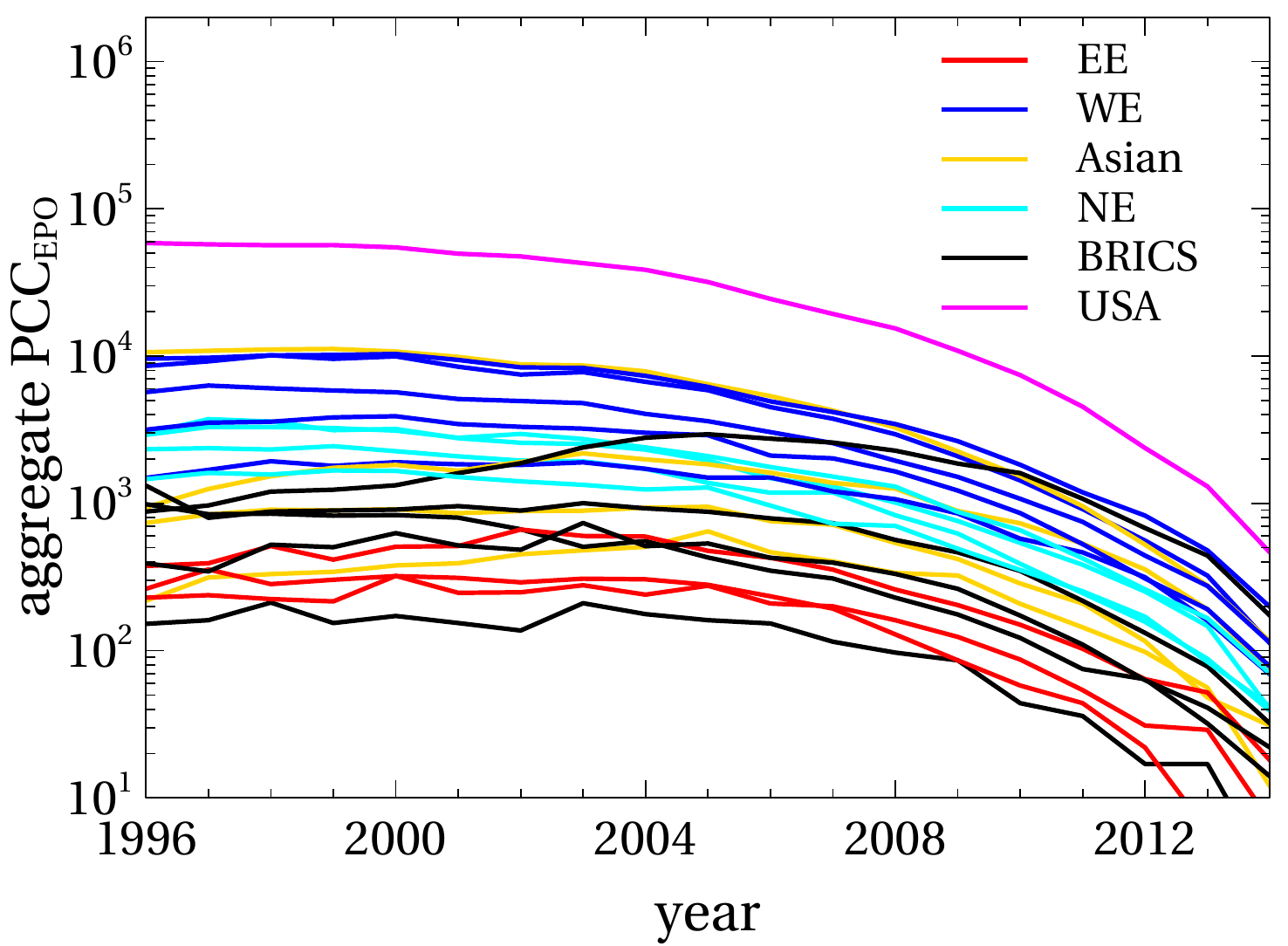}
	\includegraphics[scale=0.5]{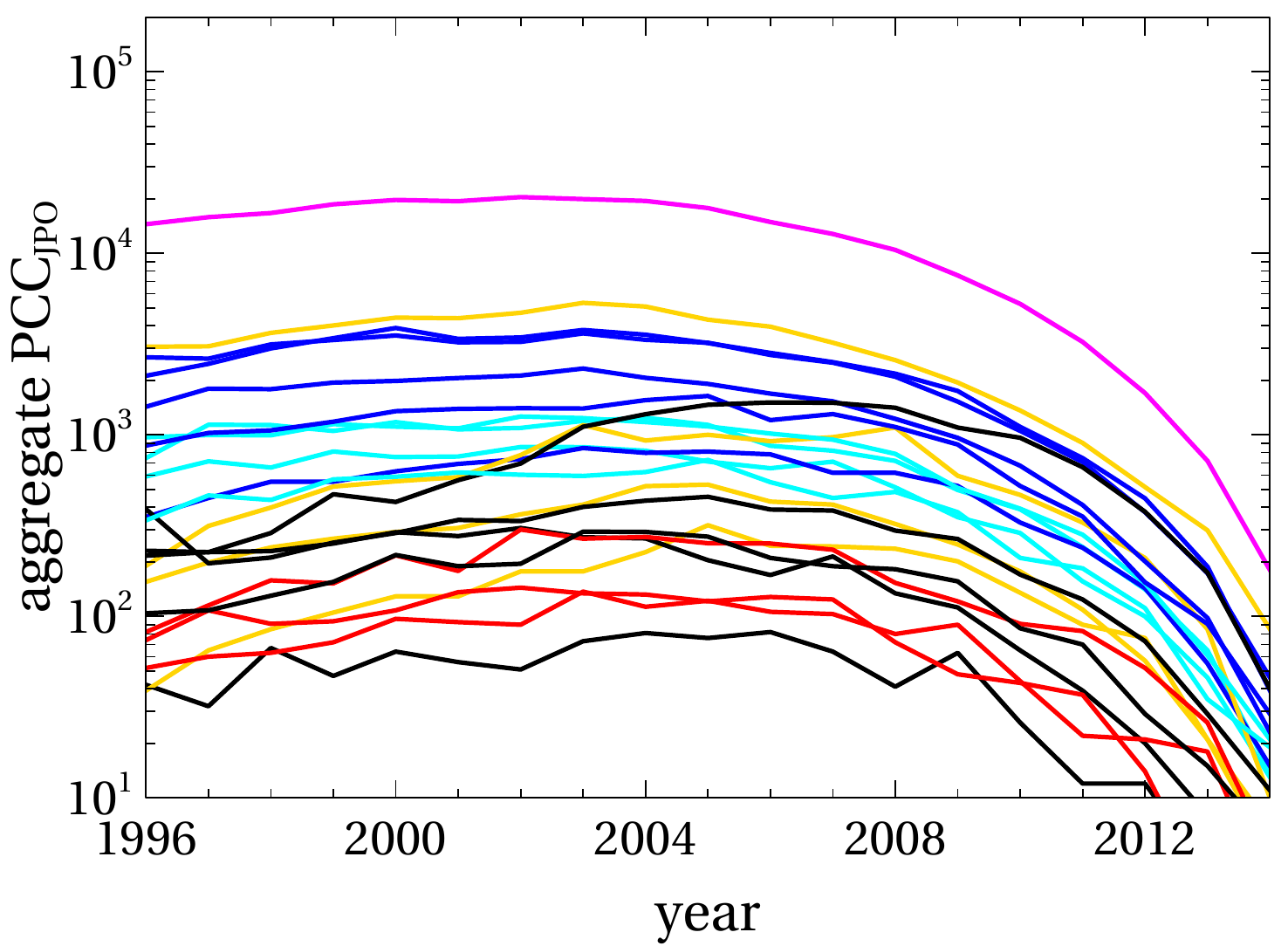}\\
	\includegraphics[scale=0.5]{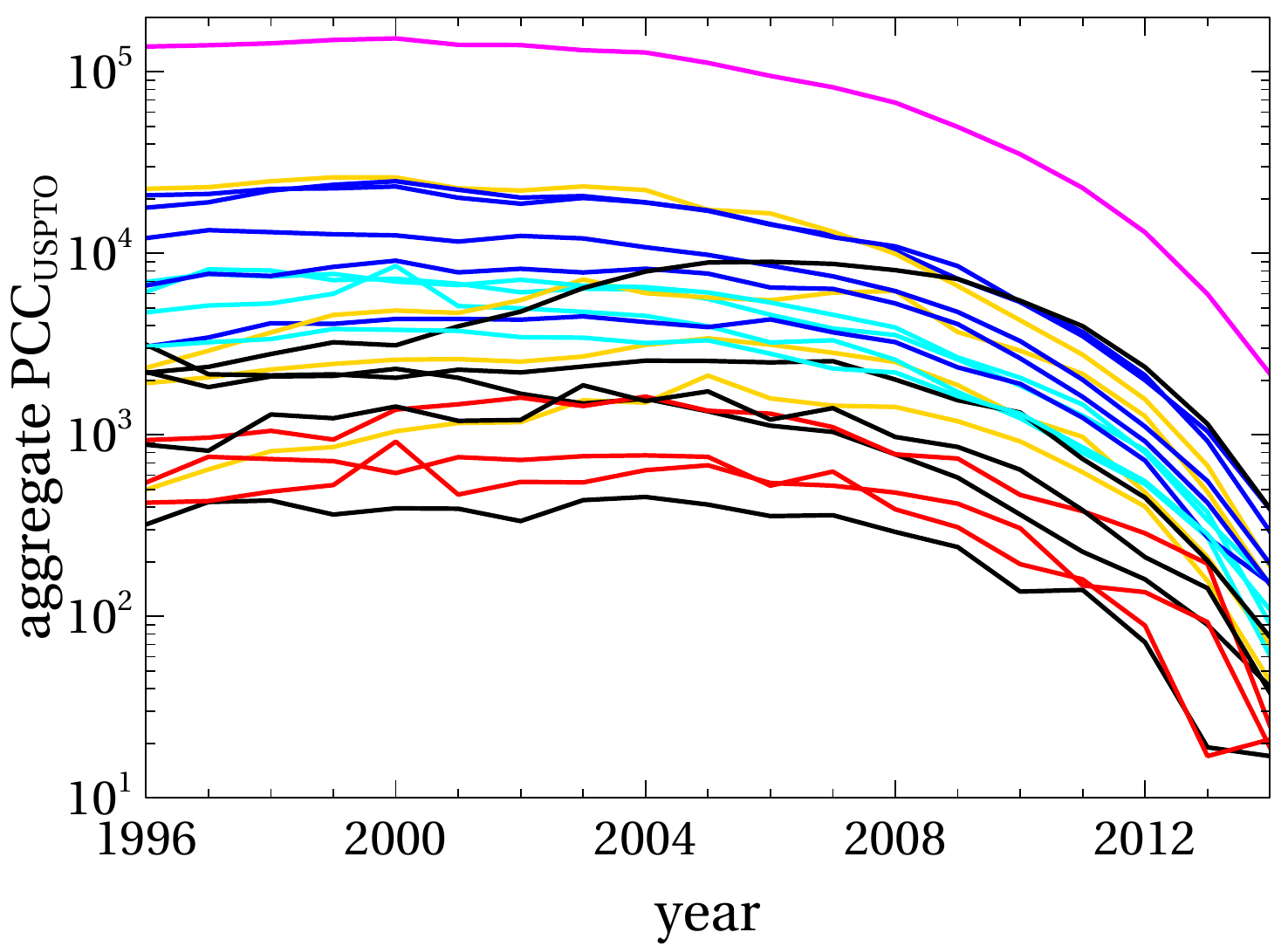}
	\includegraphics[scale=0.5]{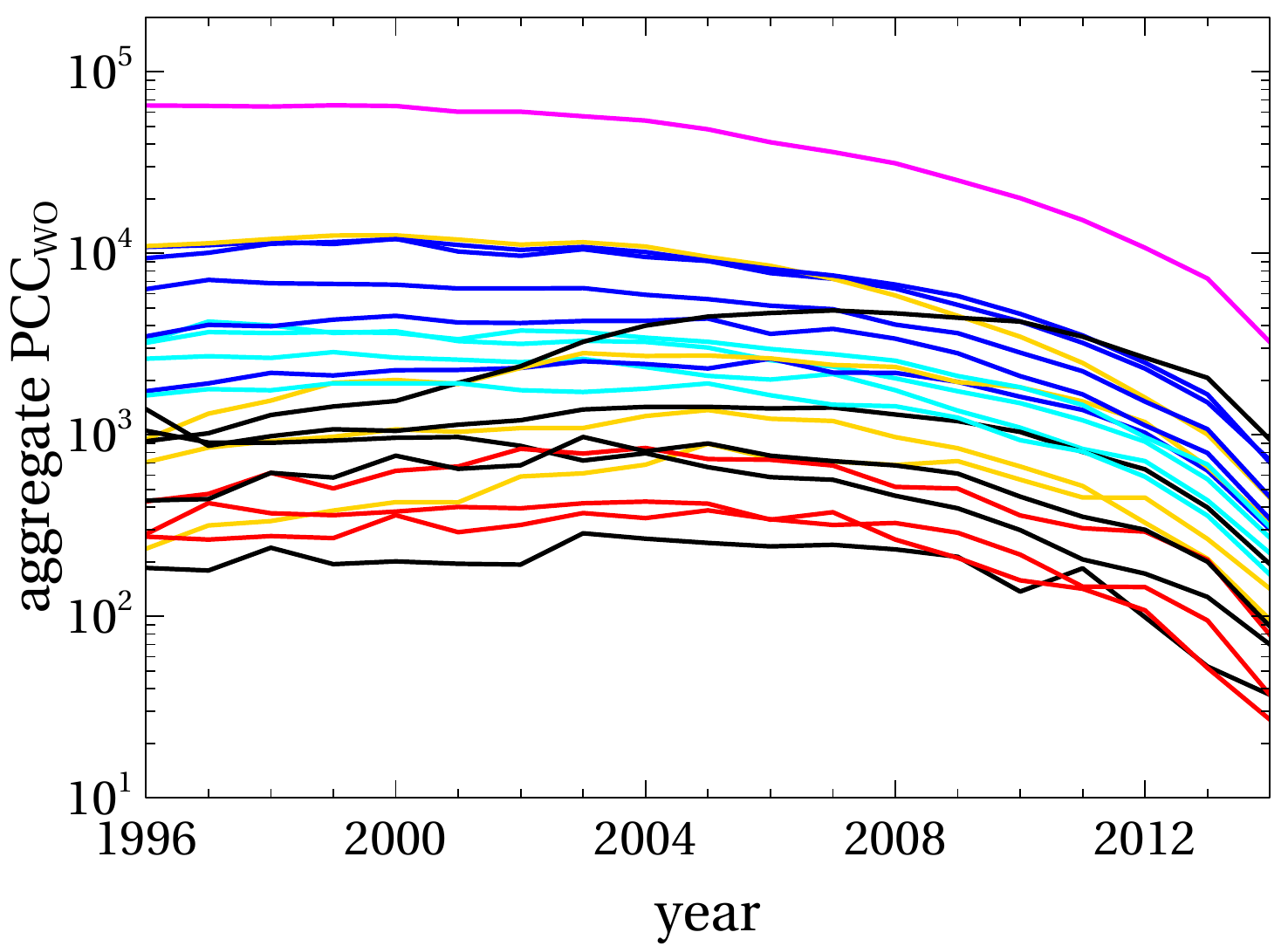}
	\caption{Time series for the count of patent citations to journal research papers (PCC), aggregated over all scientific sectors, for the main patent offices in our database: 
	EPO (top left), JPO (top right), USPTO (bottom left) and WIPO (bottom right).
	A similar qualitative behavior is observed in the plots, meaning that at this aggregation scale no particular bias toward the home scientific school emerges. 
	These plots also allow to quantify how many patent citations come from each of the patent office considered.}
	\label{fig:pat_off}
\end{figure}

\begin{figure}[h!]
	\centering
	\includegraphics[scale=0.5]{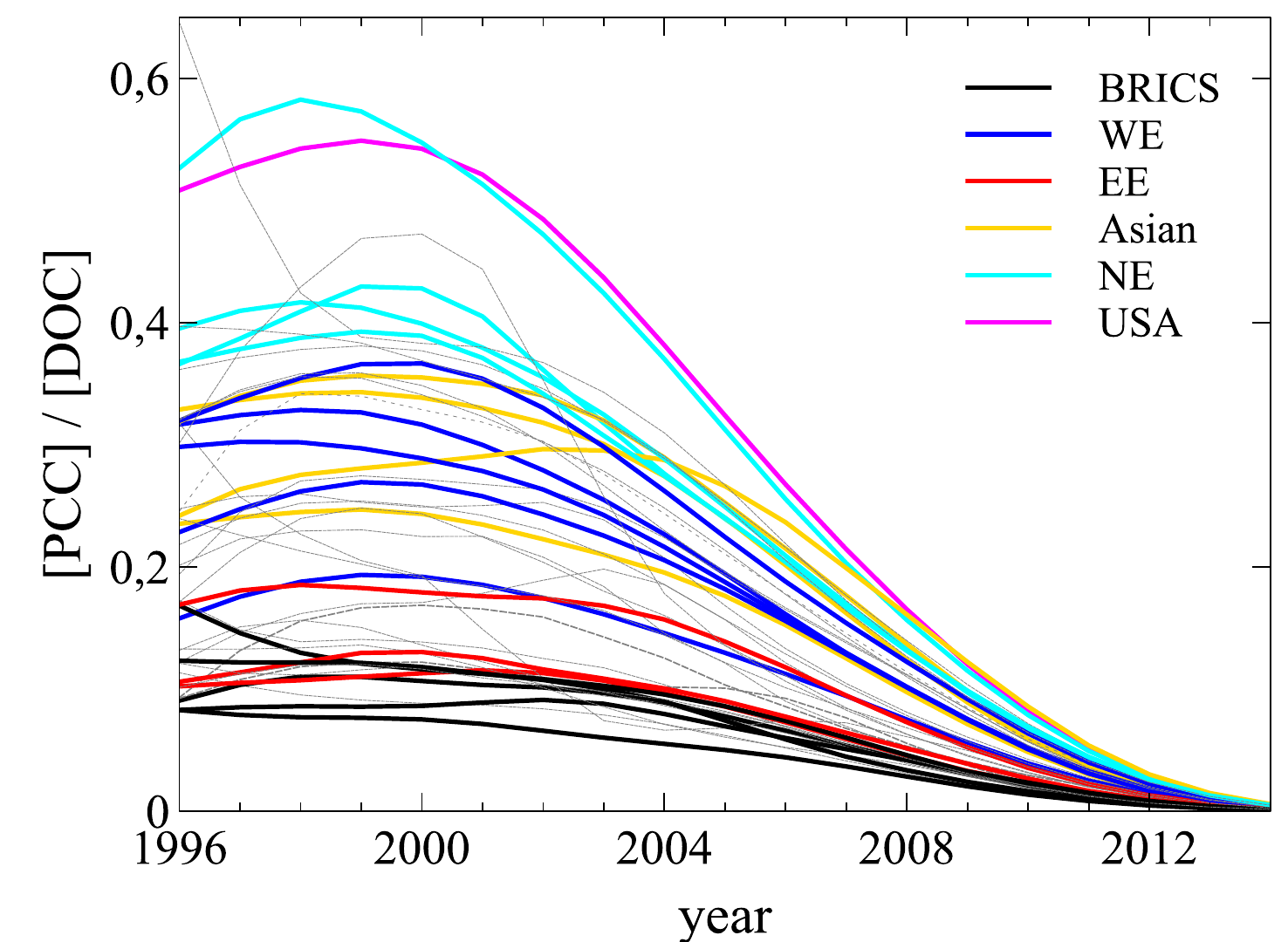}\\
	\includegraphics[scale=0.5]{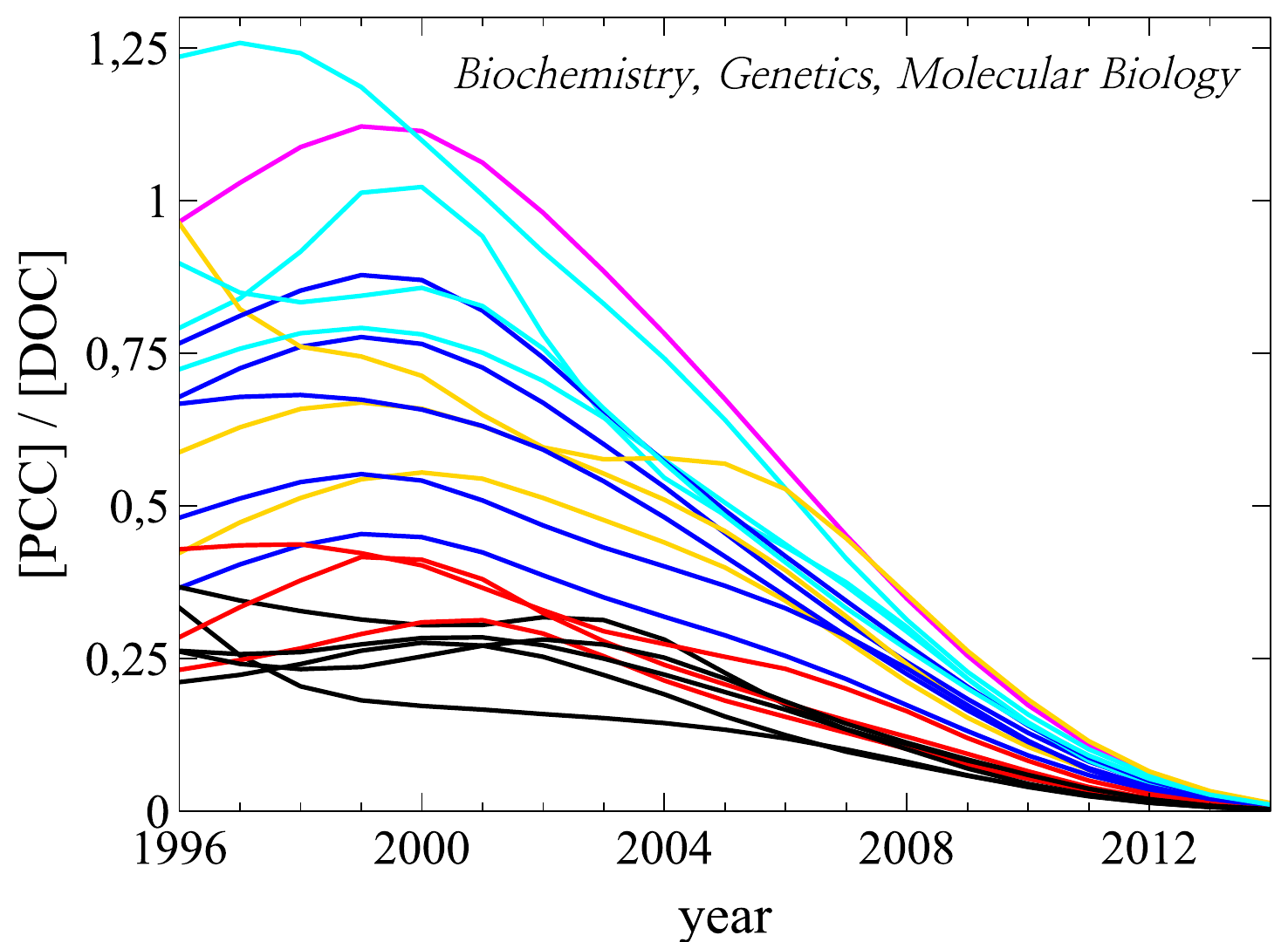}
	\includegraphics[scale=0.5]{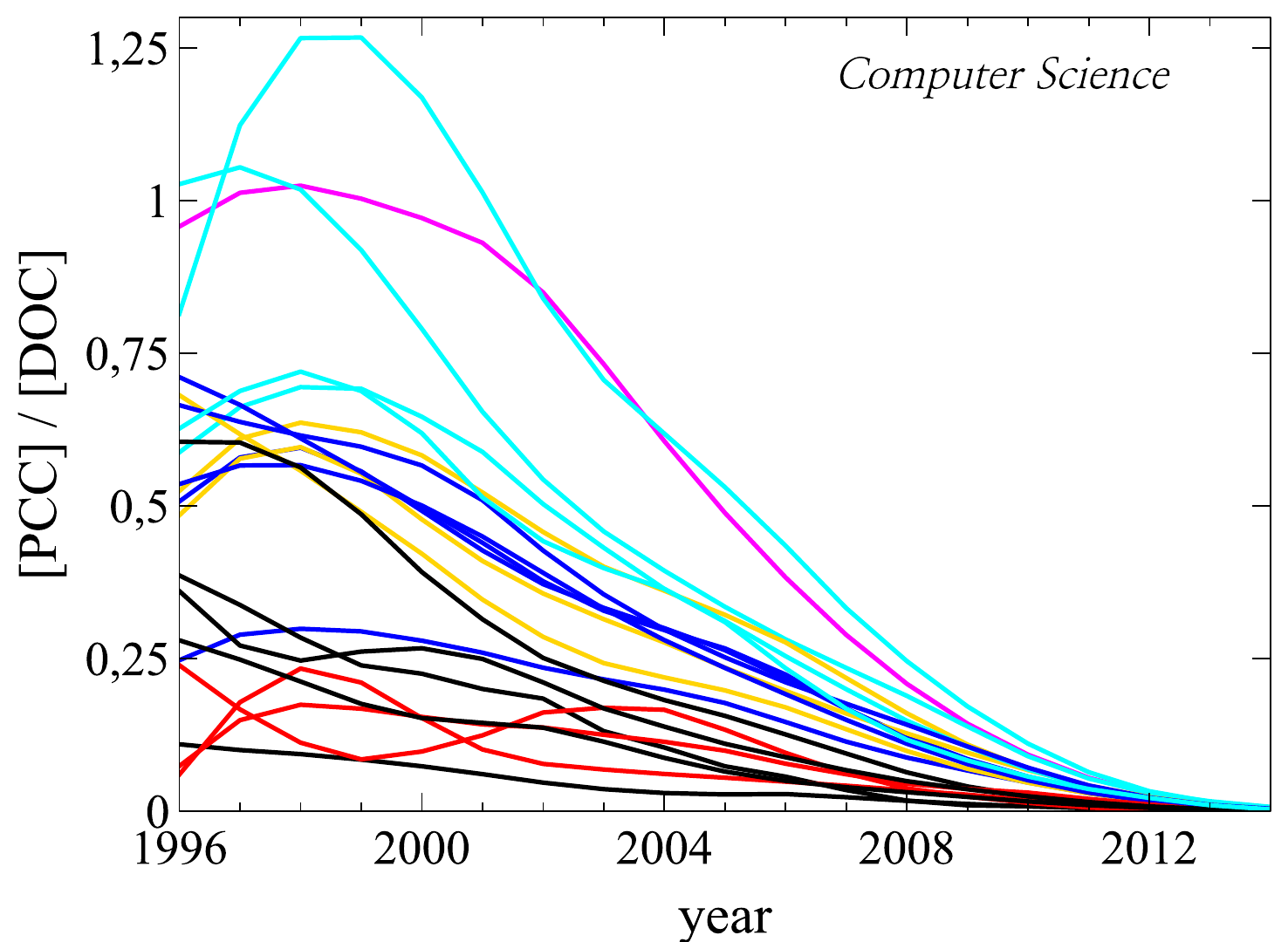}\\
	\includegraphics[scale=0.5]{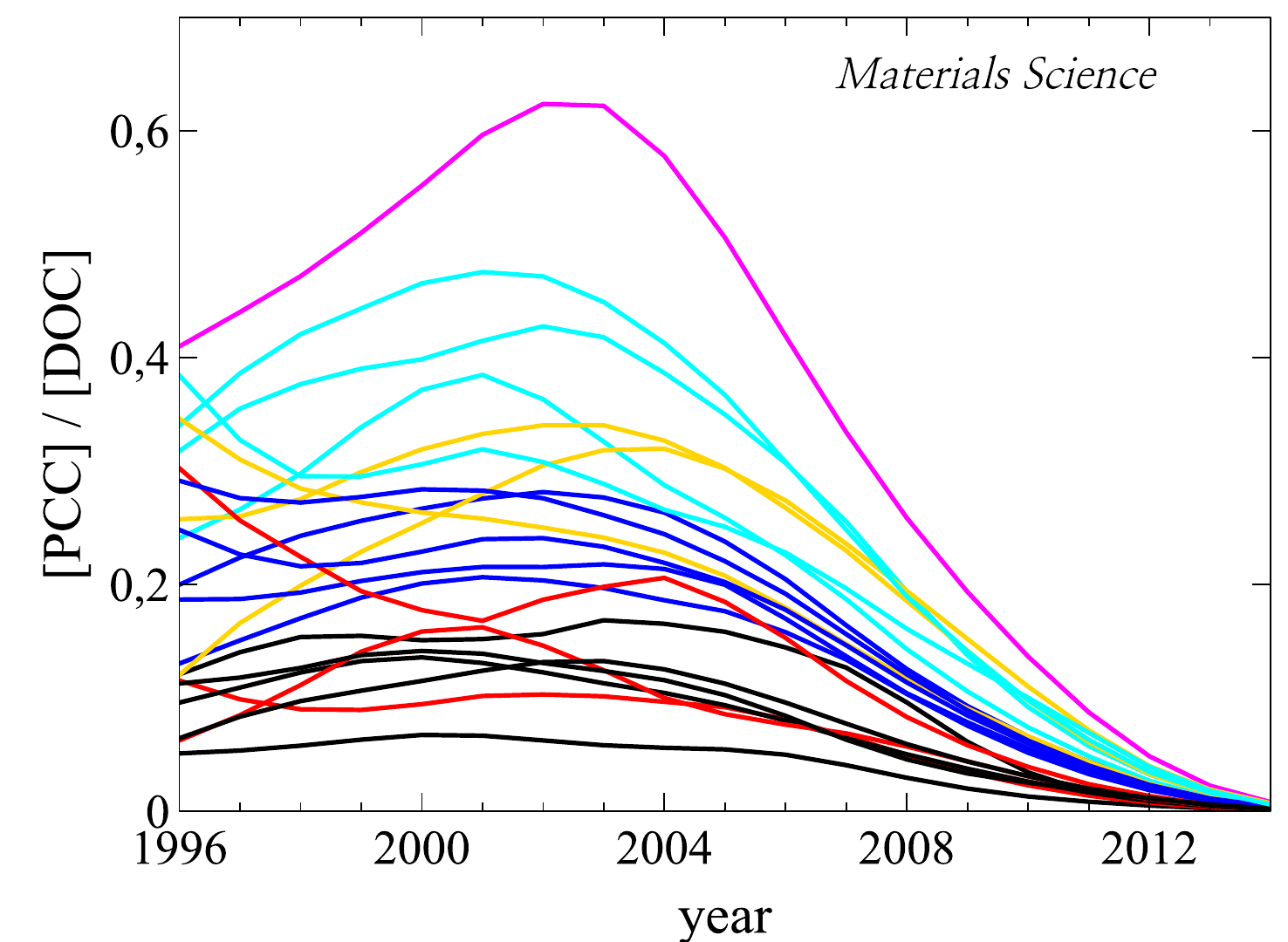}
	\includegraphics[scale=0.5]{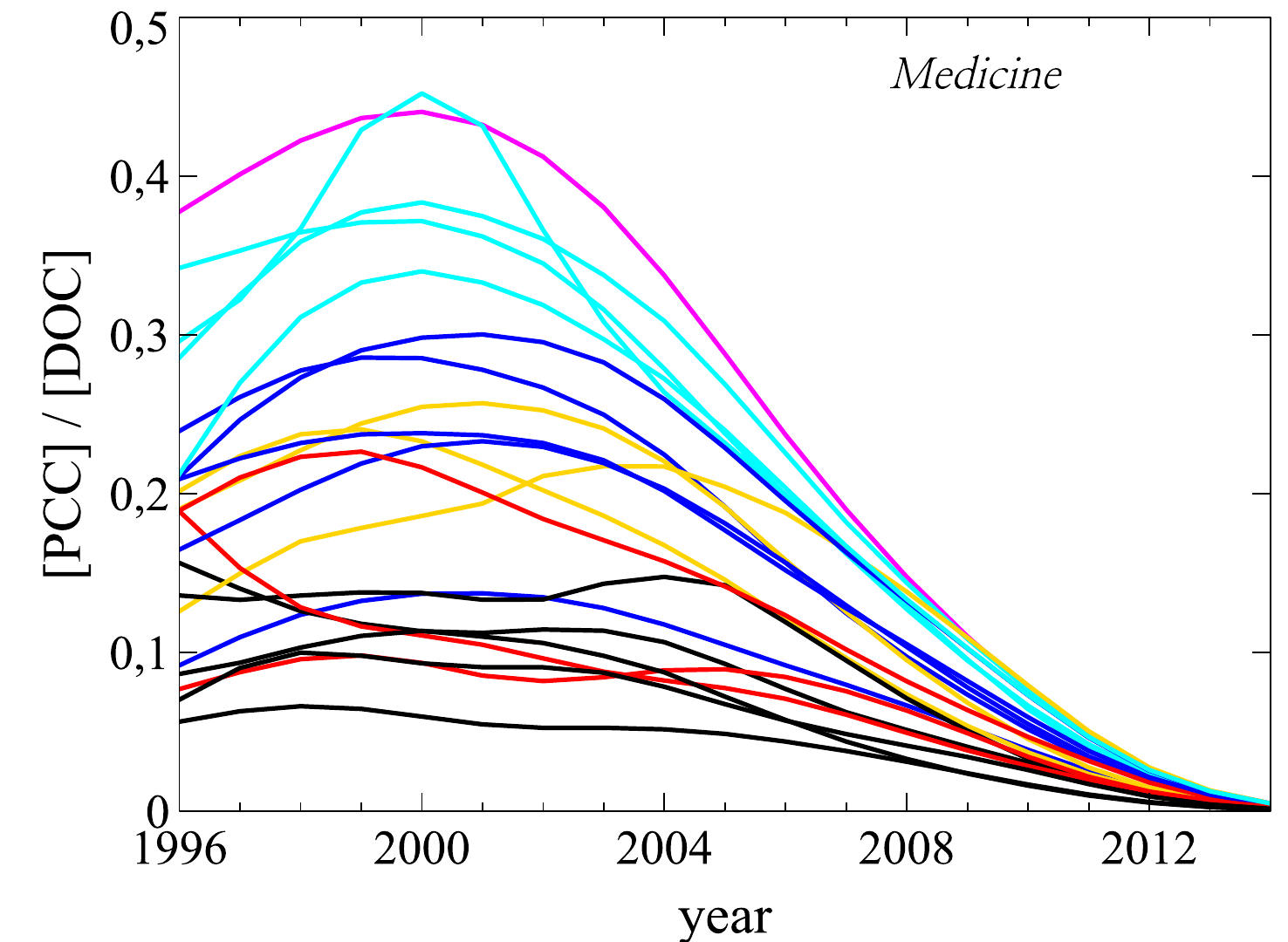}\\
	\includegraphics[scale=0.5]{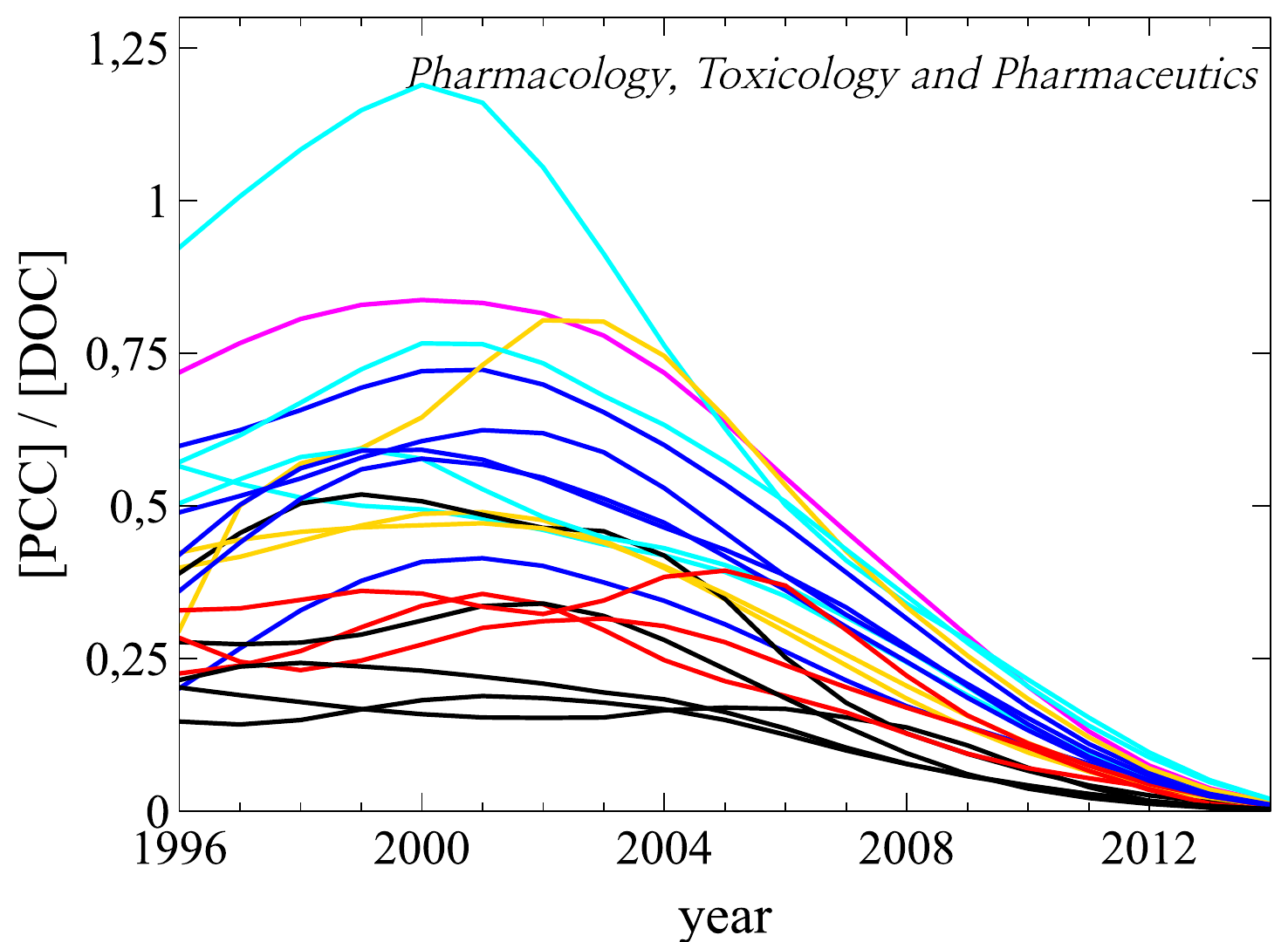}
	\includegraphics[scale=0.5]{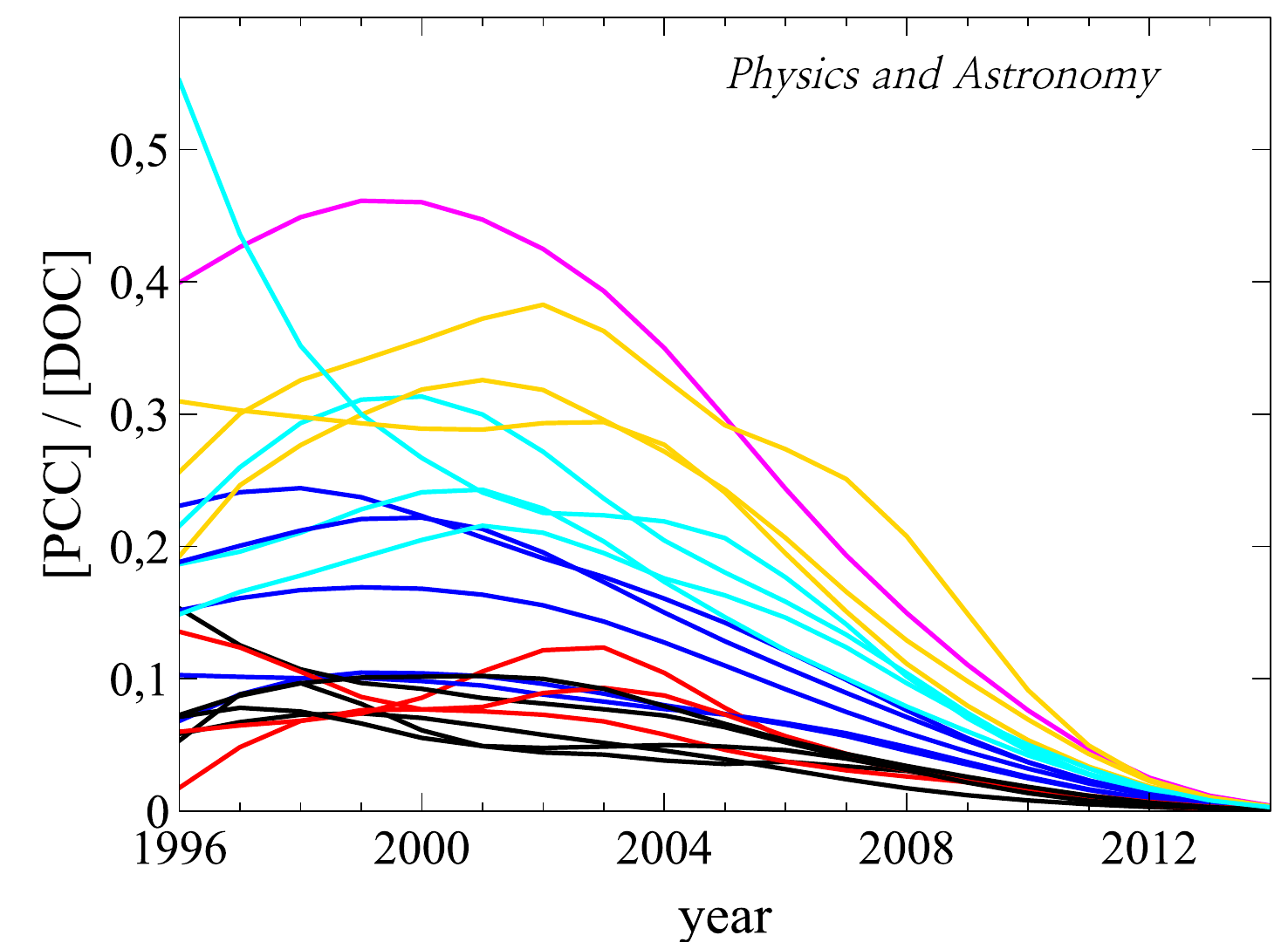}
	\caption{PCC/DOC ratios (\ie, mean number of patent citations per research article) computed on all scientific sectors (top panel) and on selected sectors: 
	{\em Biochemistry, Genetics, Molecular Biology} (upper left panel), {\em Computer Science} (upper right panel), 
	{\em Materials Science} (middle left panel), {\em Medicine} (middle right panel), {\em Pharmacology, Toxicology, Pharmaceutics} (lower left panel), {\em Physics and Astronomy} (lower right panel). 
	We see that each sector possesses its own features, however the separation between USA, Europe and BRICS is persistent.}
	\label{fig:sectors}
\end{figure}

\end{document}